\documentclass[journal]{IEEEtran}

\usepackage[noadjust]{cite}
\usepackage{amssymb} %
\usepackage{url}
\newcommand{\mvar}[1]{\ensuremath{\mathit{#1}}}
\newcommand{\mdef}[1]{\ensuremath{\mathsf{#1}}}
\newcommand\mconstr[1]{\mdef{#1}}

\newcommand{\hairspace}{\hspace{1pt}}
\newcommand{\eg}{\mbox{e.\hairspace{}g.,} }  %
\newcommand{\ie}{\mbox{i.\hairspace{}e.,} }  %

\newcommand{\etal}{\mbox{et~al.}\ }

\newcommand{\topos}{\emph{topoS}}
\newcommand{\fffuu}{\emph{{f}{f}{f}uu}}

\usepackage[dvipsnames]{xcolor}

\usepackage{setspace}

\usepackage{tabularx}
\usepackage[emulate=units,binary-units=true]{siunitx}

\usepackage[normalem]{ulem}

\usepackage{tikz}
\usetikzlibrary{shapes}
\usetikzlibrary{calc}

\usetikzlibrary{arrows}
\usetikzlibrary{arrows.meta} %
\tikzset{myptr/.style={-{Latex[scale=1.5]}}}%
\tikzset{myptrrev/.style={{Latex[scale=1.5]}-}}%
\tikzset{myptrdouble/.style={{Latex[scale=1.5]}-{Latex[scale=1.5]}}}%

\usepackage{enumitem}
\usepackage{calc}

\usepackage{fancyvrb}

\usepackage{moeptikz}

\tikzset{MyDoubleArrow/.style={double arrow, draw=black, anchor=west, align=center, text width=1em}}
\tikzset{MySingleArrow/.style={single arrow, draw=black, anchor=west, align=center, text width=1em}}
\tikzset{MySingleLeftArrow/.style={single arrow, rotate=180, draw=black, anchor=east, align=center, text width=1em}}

\tikzset{MyRoundedBox/.style={
		draw,
		rounded corners=3pt,
		inner sep=5pt,
		anchor=west,
		text width=10em,
		align=center
	}
}

\definecolor{TUMBlue}		{rgb}{0.00,0.40,0.74}	%

\definecolor{TUMWhite}		{rgb}{1.00,1.00,1.00}	%
\definecolor{TUMBlack}		{rgb}{0.00,0.00,0.00}	%

\definecolor{TUMDarkerBlue}	{rgb}{0.00,0.32,0.58}	%
\definecolor{TUMDarkBlue}	{rgb}{0.00,0.20,0.35}	%

\definecolor{TUMDarkGray}	{rgb}{0.20,0.20,0.20}	%
\definecolor{TUMMediumGray}	{rgb}{0.50,0.50,0.50}	%
\definecolor{TUMLightGray}	{rgb}{0.80,0.80,0.80}	%

\definecolor{TUMIvony}		{rgb}{0.85,0.84,0.80}	%
\definecolor{TUMOrange}		{rgb}{0.89,0.45,0.13}	%

\definecolor{TUMOrange}	{rgb}{0.98,0.73,0.00}

\definecolor{TUMGreen}		{rgb}{0.64,0.68,0.00}	%
\definecolor{TUMLightBlue}	{rgb}{0.60,0.78,0.92}	%
\definecolor{TUMLighterBlue}	{rgb}{0.39,0.63,0.78}	%

\definecolor{TUMPurple}		{rgb}{0.41,0.03,0.35}
\definecolor{TUMDarkPurple}	{rgb}{0.06,0.11,0.37}
\definecolor{TUMTurquois}	{rgb}{0.00,0.47,0.54}
\definecolor{TUMDarkGreen}	{rgb}{0.00,0.49,0.19}
\definecolor{TUMDarkerGreen}	{rgb}{0.40,0.60,0.11}
\definecolor{TUMYellow}		{rgb}{1.00,0.86,0.00}
\definecolor{TUMDarkYellow}	{rgb}{0.98,0.73,0.00}
\definecolor{TUMLightRed}	{rgb}{0.84,0.30,0.07}
\definecolor{TUMRed}		{rgb}{0.77,0.03,0.11}
\definecolor{TUMDarkRed}	{rgb}{0.61,0.05,0.09}

\newcommand{\diffadd}[1]{\textcolor{TUMDarkGreen}{\textbf{#1}}}
\newcommand{\diffdel}[1]{\textcolor{TUMRed}{\textit{#1}}}

\newcommand{\dockerbr}[0]{dbr}
\newcommand{\ippostfix}[0]{} %

\begin{document}

\title{Agile Network Access Control in the Container Age}

\author{Cornelius~Diekmann, %
        Johannes~Naab, %
        Andreas Korsten,
        and~Georg~Carle%

\thanks{C.\ Diekmann, J.\ Naab, A.\ Korsten and G.\ Carle are with the
Department of Informatics, Technical University of Munich (TUM), Garching bei
M\"{u}nchen 85748, Germany (e-mail: $\lbrace$diekmann, naab, korsten,
carle$\rbrace$@net.in.tum.de).}%

\thanks{Manuscript received ???? ??, 201?; revised ?????? ??, 20??.}}

\markboth{TNSM preprint}%
{Shell \MakeLowercase{\textit{et al.}}: Bare Demo of IEEEtran.cls for IEEE Journals}
\maketitle

\begin{abstract}
	Linux Containers, such as those managed by Docker, are an increasingly popular way to package and deploy complex applications.
	However, the fundamental security primitive of network access control for a distributed microservice deployment is often ignored or left to the network operations team.
	High-level application-specific security requirements are not appropriately enforced by low-level network access control lists.
	Apart from coarse-grained separation of virtual networks, Docker neither supports the application developer to specify nor the network operators to enforce fine-grained network access control between containers.

	In a fictional story, we follow DevOp engineer Alice through the lifecycle of a web application. 
	From the initial design and software engineering through network operations and automation, we show the task expected of Alice and propose tool-support to help.
	As a full-stack DevOp, Alice is involved in high-level design decisions as well as low-level network troubleshooting. 
	Focusing on network access control, we demonstrate shortcomings in today's policy management and sketch a tool-supported solution. 
	We survey related academic work and show that many existing tools fail to bridge between the different levels of abstractions a full-stack engineer is operating on.
	
	Our toolset is formally verified using Isabell/HOL and is available as Open Source.
\end{abstract}


\IEEEpeerreviewmaketitle

\section{Introduction}
\label{sec:introduction}
\IEEEPARstart{N}{etwork-level}
 access control is a fundamental security mechanism, not only in traditional networks, but also in distributed applications, clouds, and microservice architectures. %
Unfortunately, configuring network-level access control still is a challenging, manual, and thus error-prone task~\cite{fwviz2012,fireman2006,ZhangAlShaer2007flip}. %
It is a known and unsolved problem for over a decade that ``corporate firewalls are often enforcing poorly written rule sets''~\cite{firwallerr2004}. 
Also, ``access list conflicts dominate the misconfiguration errors made by administrators''~\cite{netsecconflicts}. 
A recent study confirms that this problem persists as a ``majority of administrators stated misconfiguration as the most common cause of failure''~\cite{sherry2012making}.
In addition, not only is implementing a policy error-prone, but also designing it is challenging, even for experienced administrators~\cite{diekmann2014forte}.

In this article, we tell a fictional story about administrator Alice. 
Alice is responsible for designing and operating a distributed web application. 
She uses Linux containers managed by Docker~\cite{dockerweb}. 
The story covers both the design phase and operations. 
Alice is not responsible for the application logic, but she is responsible for helping the application scale and for network security. 
In modern terminology, Alice can be called a DevOp or SRE (Site Reliability Engineer)~\cite{google2016sre}. 
Alice knows that when using Docker it is best practice to decrease the attack surface by limiting container networking~\cite{docker2015sec} and our story primarily focuses on network-level access control.

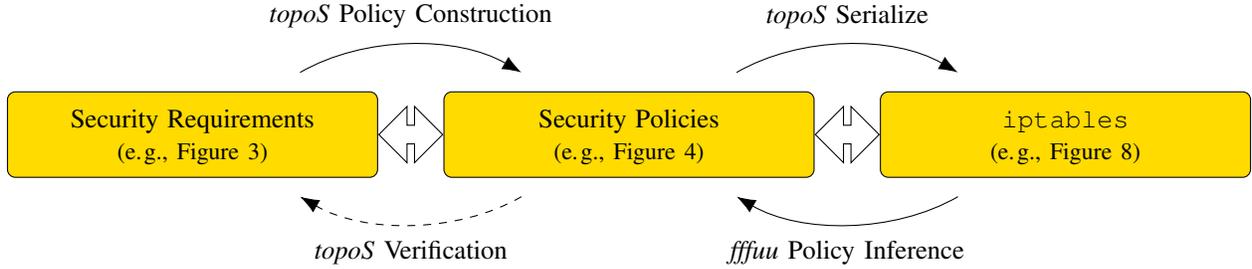
\begin{figure*}[!bht]
	\centering
	 \begin{tikzpicture}
	  \node [MyRoundedBox, fill=TUMYellow, text width=13em](sinvar) at (0,0) {\strut{}Security Requirements\\\small{(\eg Figure~\ref{fig:runexsecinvars})}};
	  \node [MyDoubleArrow](arr1) at (sinvar.east) {};
	  \node [MyRoundedBox, fill=TUMYellow, text width=13em](policy) at (arr1.east) {\strut{}Security Policies\\\small{(\eg Figure~\ref{fig:secpol})}};
	  \node [MyDoubleArrow](arr2) at (policy.east) {};
	  \node [MyRoundedBox, fill=TUMYellow, text width=13em](mechanism) at (arr2.east) {\strut{}\texttt{iptables}\\\small{(\eg Figure~\ref{fig:dockerfirewall:tuned})}};
	 
	  \path[draw,myptr,shorten >=0.5cm,shorten <=0.5cm] 
	  (sinvar) to[bend left]    node[anchor=south, yshift=1ex] {\topos{} Policy Construction} (policy);
	
	  \path[draw,myptr,shorten >=0.5cm,shorten <=0.5cm] 
	  (policy) to[bend left]    node[anchor=south, yshift=1ex] {\topos{} Serialize} (mechanism);
	
	  \path[draw,myptr,shorten >=0.5cm,shorten <=0.5cm] 
	  (mechanism) to[bend left]    node[anchor=north, yshift=-1ex] {\fffuu{} Policy Inference} (policy);
	
	  \path[draw,dashed,myptr,shorten >=0.5cm,shorten <=0.5cm] 
	  (policy) to[bend left]    node[anchor=north, yshift=-1ex] {\topos{} Verification} (sinvar);
  	\end{tikzpicture}%
	\caption{Overview of the Tools \topos{} and \fffuu{} Bridging Between High Abstraction Levels (left) and Low-Level Details (right)}\label{fig:toolsoverview}
\end{figure*}

We present two tools which help Alice in various situations. 
First, \topos{}~\cite{diekmann2015topos} is a constructive, top-down greenfield approach for network security management.
\topos{} translates high-level security goals to Linux iptables firewall configurations, which can be installed on a Docker host. %
The automatic translation steps prevent manual translation errors. 
Furthermore, \topos{} visualizes the results of all translation steps to help Alice uncover specification errors and allow low-level tuning. 
The second tool \fffuu{}~\cite{diekmann2016networking} is complementary to \topos{}: 
\fffuu{} digests existing iptables rules and visualizes their filtering behavior. 
This direction is particularly challenging due to the vast amount of modules and low-level features which can be used in iptables. 
We chose Docker as a particularly challenging environment as Docker in its early days was known to ``thrash and [destroy] you\,[\textit{sic}] iptables rules, network interfaces hierarchy and routing tables''~\cite{boycottdocker}. 
An overview of our tools is sketched in Figure~\ref{fig:toolsoverview}.

\topos{} \& \fffuu{} are not specific to our case study. 
Both tools are formally verified~\cite{Network_Security_Policy_Verification-AFP,Iptables_Semantics-AFP} with \mbox{Isabelle/HOL}~\cite{isabelle2016}. 
Isabelle is an LCF-style interactive theorem prover; the correctness of all proven facts is based on the correctness of a small mathematical inference kernel. 
This architecture is very robust and not a single bug which practically affects a user's proof emerged since nearly 20 years. 
As Isabelle is an interactive proof assistant---in contrast to automated theorem provers---proofs in Isabelle often require a significant amount of work. 
In return, Isabelle provides a high level of confidence about the correctness of the proven facts. 
Our tools \topos{} and \fffuu{} took several years to be developed and verified. 
As a result, we contribute formally verified tools which are proven correct for all inputs, can run stand-alone without Isabelle, do not require any manual proof from Alice, nor expose overformalization. 
Their core functionality can also be reused as a library in further projects. 
In this article, we will not present the formal background~\cite{diekmann2014forte,diekmann2014EPTCS,diekmann2016networking,Network_Security_Policy_Verification-AFP,Iptables_Semantics-AFP}, instead, we demonstrate applicability from an operator's point of view; not requiring a single formula. 

It would have been possible to carry out the development in a different interactive theorem prover, for example Coq~\cite{coqmanual}. 
In contrast, tools such as model checkers, automated theorem provers (atp), or smt solvers are not sufficient for this task. 
Traditional model checkers are unsuitable since one cannot simply exhaust all the state space of our model (for example, our model includes an arbitrary function to model an oracle for iptables match conditions, thus also supporting match conditions which are not even developed yet~\cite{diekmann2015fm}). 
In addition, atps and smt solvers usually fail or time out on the complicated proof obligations. 
Isabelle employs many state-of-the-art atps and smt solvers to help automatically discharging proof obligations, but very often, the core ideas of a proof or ingenious helping lemmas are discovered by a human.

This article is partly based on our previously published paper~\cite{diekmann2015topos}. 
Our previous publication discusses the design phase and provides a formally-verified method to translate security requirements to a security policy~(\mbox{\S\hairspace\ref{sec:sdnnfv:topos}}). 
In this article, we use those initial results and show its integration with Docker~(\mbox{\S\hairspace\ref{sec:deploytodocker}}). 
Additionally, we use \fffuu{} to verify the low-level iptables rules, which is crucial for the non-trivial interaction of the Docker-generated rules and our \topos{}-generated \texttt{iptables} rules. 
Ultimately, this enables the usage of our tools not only in a clean slate design, but also in non-trivial operations (\mbox{\S\hairspace\ref{sec:operations}}). 
To the best of our knowledge, this is the first time that formally-verified tools are presented to help operators bridging the gaps between the abstraction level of Figure~\ref{fig:toolsoverview} in both directions. 

\noindent Our key contribution are:
\begin{itemize}
	\item We apply the formally verified tools \topos{} and \fffuu{} for network access control management in container cloud environments.
	\item We investigate how a network operator can create a formally verified firewall ruleset based on high level security goals.
	\item We provide a method to easily understand feedback on changes to the low level firewall rules.
	\item We review the related work to show that this is the first time, that a comprehensive solution to map from security policies to enforcement device implementation and back has been provided.
\end{itemize}

The rest of this article is structured as follows. 
We tell how Alice is designing the network in Section~\ref{sec:sdnnfv:topos} (based on previous publication~\cite{diekmann2015topos}). 
Alice deploys her setup with Docker in Section~\ref{sec:deploytodocker}. 
In Section~\ref{sec:operations}, the service goes live and we track Alice as operator. 
We present related Docker work in Section~\ref{sec:relateddockerwork} and related academic work in Section~\ref{sec:sdnnfv:related} (extending on previous publication~\cite{diekmann2015topos}). 
Finally, we discuss at the example of the fictional story in Section~\ref{sec:relatedcomparison} how the two tools presented enhance the state-of-the-art. 

\section{Designing the network with \topos{}}
\label{sec:sdnnfv:topos}
The security requirements of distributed applications depend on the usage scenario.
Alice utilizes the tool \topos{} to configure and design the network architecture according to the needs of her specific web application. 
Alice specifies the high-level security requirements and \topos{} synthesizes the low-level iptables rules for her.
\topos{} suggests the following workflow: 
\begin{enumerate}[label=\Alph*.]
	\item\label{topos:step:a} Formalize high-level security goals%
	\begin{enumerate}%
		\item\label{topos:step:a:a} Categorize security goals
		\item\label{topos:step:a:b} Add scenario-specific knowledge
		\item\label{topos:step:a:c} Auto-complete information ($\mathbf{\star}$)
	\end{enumerate}
	\item\label{topos:step:b} Construct security policy ($\mathbf{\star}$)
	\item\label{topos:step:c} Construct stateful policy ($\mathbf{\star}$)
	\item\label{topos:step:d} Serialize iptables configurations ($\mathbf{\star}$)
\end{enumerate}

All steps annotated with an asterisk are automated by \topos{}.
As the ($\mathbf{\star}$)-steps illustrate, once the security goals are specified, the process is completely automatic.
Between the automated steps, Alice may manually refine the intermediate result. 
\topos{} supports re-verification of the manual refinement to prevent the introduction of human errors. 

The automated intermediate ($\mathbf{\star}$)-steps are proven correct for all inputs~\cite{Network_Security_Policy_Verification-AFP}. 
The proofs are machine-verified with \mbox{Isabelle/HOL}~\cite{isabelle2016}. 
Thus, it is guaranteed that \topos{} performs correct transformations.
As a side note, since the intermediate transformations are proven correct once and for all for all inputs, Alice does not need to prove anything manually. %
The final serialization of iptables configurations is not verified since it is merely syntactic rewriting of the result of the previous step. %
Alice will also later change this rewriting slightly to better accommodate for her Docker environment. 
To prevent errors in this ad-hoc low-level step, Alice later verifies her resulting configuration with \fffuu{}. 

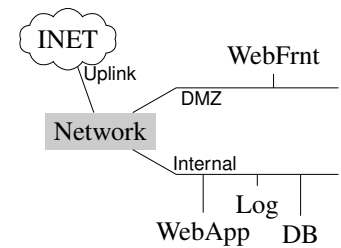
\begin{figure}[htb]
\centering
\resizebox{0.5\linewidth}{!}{%
		    \begin{Large}
		    \begin{tikzpicture}
			   	\node[anchor=center,align=center,text width=8.5em,cloud, draw,cloud puffs=10,cloud puff arc=120, aspect=2, inner sep=-3em,outer sep=0] (inet) at (0,-.3) { INET };
			   	\node[anchor=center,rectangle, fill=TUMLightGray] (networkbox) at (.6,-2) {Network};
			   	
			   	\node[anchor=center] (webfrnt) at (3.8,-.6) { WebFrnt };
			   	\node[anchor=center] (webapp) at (2.5,-3.9) { WebApp };
			   	\node[anchor=center] (log) at (3.5,-3.4) { Log };
			   	\node[anchor=center] (db) at (4.3,-3.9) { DB };
		
			    \draw (inet) node[anchor=center,label={[below,xshift=3.5ex,yshift=-2.3ex]\normalsize{\textsf{Uplink}}}] {}--(networkbox);
			    \draw (networkbox)--(2,-1.2)node[anchor=center,label={[below,xshift=2.1ex,yshift=-.3ex]\normalsize{\textsf{DMZ}}}] {}--(5,-1.2);
			    \draw (networkbox)--(2,-2.8)node[anchor=center,label={[above,xshift=2.4ex,yshift=-1.2ex]\normalsize{\textsf{Internal}}}] {}--(5,-2.8);
			    
			    \draw (3.8,-1.2)--(webfrnt);
			    \draw (2.5,-2.8)--(webapp);
			    \draw (3.5,-2.8)--(log);
			    \draw (4.3,-2.8)--(db);
		    \end{tikzpicture}
		    \end{Large}}
			\caption{Network Schematic}\label{fig:netschematic}
\end{figure}

Now we consider the actual web application. 
The scenario was chosen because it has been used previously~\cite{diekmann2015topos}, is minimal and comprehensible for an article, but also realistic and features many important aspects.
Alice schematically illustrates the overall architecture in Figure~\ref{fig:netschematic}. 
The grey box represents the Docker network. 
The setup hosts a news aggregation web application, accessible from the Internet (\emph{INET}). 
It consists of a web application backend server (\mbox{$\mvar{WebApp}$}) and a frontend server (\mbox{$\mvar{WebFrnt}$}). 
The $\mvar{Web\-App}$ is connected to a database ($\mvar{DB}$) and actively retrieves data from the Internet. 
All servers send their logging data to a central, protected log server ($\mvar{Log}$). 

Alice implements the network-related aspects of the scenario with different protocols and technologies.
The custom backend, the $\mvar{WebApp}$ is written in \texttt{python}. %
The $\mvar{WebFrnt}$ runs \texttt{lighttpd}. 
It serves static web pages directly and retrieves dynamic websites from the $\mvar{WebApp}$. %
All components send their \texttt{syslog} messages via UDP (RFC~5426~\cite{rfc5426}) to $\mvar{Log}$. 
Since the implementation details are irrelevant, we prototyped the setup and checked connectivity with \texttt{busybox} container images. 

For details on the architecture and working principles of \topos{} we refer to the original publication \cite{diekmann2014forte}.

\subsection{Formalizing High-Level Security Goals}
\label{sub:securitygoals}
Formalizing the security goals, \ie step~\ref{topos:step:a} in the process of using \topos{}, is the most crucial and manual part. 
First, Alice collects the entities in her setup: $\mvar{INET}$, \mbox{$\mvar{WebApp}$}, \mbox{$\mvar{WebFrnt}$}, $\mvar{DB}$, and $\mvar{Log}$. 
Now, \topos{} provides a modular, attribute-based language~\cite{diekmann2014forte} to specify the security requirements. 
\topos{} comes with a pre-defined library of security invariant templates as listed in Table~\ref{tab:securityinvarianttemplates}.

\begin{table}[htb]
	\small
	\noindent\begin{tabularx}{\columnwidth}{lX}
		Name & Description \\
		\hline
		Simple BLP & Simplified Bell-LaPadula \\
		Bell-LaPadula & Label-based Information Flow Security with trusted entities \\
		ACL & Simple ACLs (Access Control Lists) \\
		Comm. With & White-listing transitive ACLs \\
		Not Comm. With & Black-listing transitive ACLs \\
		Dependability & Limit dependence on certain hosts \\
		Domain Hierarchy & Hierarchical control structures \\
		NoRefl & Allow/deny reflexive flows. Can lift symbolic policy identifiers to role names (e.g., symbolic host name corresponds to an IP range.) \\
		NonInterference & Transitive non-interference properties \\
		PolEnforcePoint & Central application-level policy enforcement point. Master/Slave relationships. \\
		Sink & Information sink. Hosts (or host groups) must not publish any information \\
		Subnets & Collaborating, protected host groups \\
		SubnetsInGW & Simple, collaborating, protected or accessible host groups \\
		Simple Tainting & Simplified label-based Privacy \\
		Tainting & Label-based Privacy with untainting \\
	\end{tabularx}
	\caption{Security Invariant Templates defined by \topos}
	\label{tab:securityinvarianttemplates}
\end{table}

A template formalizes generic, scenario-independent aspects of a security goal and must be formally defined in \topos{} using \mbox{Isabelle/HOL}. 
Alice only instantiates those templates by adding scenario-specific information. 
She does so by assigning attributes to entities. 
\topos{} does not require Alice to assign attributes to all entities. 
Alice must only provide all security-relevant information and \topos{} auto-completes the missing values with provably secure default values~\cite{diekmann2014forte}. 
For this scenario, Alice instantiates four invariant templates to define her security goals. 
We now explain the invariants in the language of \topos{}. 
Figure~\ref{fig:runexsecinvars} shows the final specification Alice writes. %

\begin{figure}[!htb]
	\centering
	\fbox{
			\resizebox{.8\linewidth}{!}{%
				\begin{minipage}{\widthof{Bell LaPadula $\lbrace\mvar{WebApp} \mapsto \mdef{declassify} \ (\mdef{trusted})\rbrace$}}
\begin{flushleft}
	\smallskip
	Subnets $\lbrace \mvar{DB} \mapsto \mconstr{internal}$, \newline
	\phantom{Subnets $\lbrace$}$\mvar{Log} \mapsto \mconstr{internal}$, \newline
	\phantom{Subnets $\lbrace$}$\mvar{WebApp} \mapsto \mconstr{internal}$, \newline
	\phantom{Subnets $\lbrace$}$\mvar{WebFrnt} \mapsto \mconstr{DMZ} \rbrace$
	
	\medskip
	
	Sink $\lbrace \mvar{Log} \mapsto \mconstr{sink} \rbrace$
	
	\medskip
	
	Bell LaPadula $\lbrace \mvar{DB} \mapsto \mdef{confidential},\ $\newline
	\phantom{Bell LaPadula $\lbrace$}$ \mvar{Log} \mapsto \mdef{confidential},\ $\newline
	\phantom{Bell LaPadula $\lbrace$}$\mvar{WebApp} \mapsto \mdef{declassify} \rbrace$
	
	\medskip
	
	ACL $\lbrace \mvar{DB} \mapsto \text{Access\ allowed\ by}: \mvar{WebApp}
	\rbrace$\\
	\smallskip
\end{flushleft} 				\end{minipage}
			}
	}
	\caption{Security Invariants}\label{fig:runexsecinvars}
\end{figure}

\begin{enumerate}
\item First, as illustrated in Figure~\ref{fig:netschematic}, $\mvar{DB}$, $\mvar{Log}$, and $\mvar{WebApp}$ are considered internal hosts. 
Alice uses a template called \emph{Subnets}. 
She labels internal hosts with the $\mconstr{internal}$ attribute. 
The $\mvar{WebFrnt}$ must be accessible from outside, it is a classical $\mconstr{DMZ}$ member and labeled accordingly. %
\item Next, Alice wants to ensure that logging data must not leave the log server. 
Therefore, using a template called \emph{Sink}, she classifies $\mvar{Log}$ as information sink.

\item Using a template called \emph{Bell LaPadula}, Alice specifies that $\mvar{DB}$ contains $\mconstr{confidential}$ information. 
Since it sends its log data to the log server, she labels $\mvar{Log}$ as $\mconstr{confidential}$. 
Finally, the $\mvar{Web\-App}$ is allowed to retrieve data from the $\mvar{DB}$ and to publish it to the $\mvar{WebFrnt}$. 
Therefore, the $\mvar{Web\-App}$ is trusted and allowed to $\mdef{declassify}$ data.

\item A traditional access control list, using the \emph{ACL} template, specifies that only $\mvar{Web\-App}$ may access the $\mvar{DB}$. 
\end{enumerate}

This is all the information \topos{} needs to operate. 
Based on this, \topos{} can now compute a security policy as access control graph, shown in Figure~\ref{fig:secpol}. 
This policy is much more fine-grained than the simple DMZ architecture Alice initially drafted in Figure~\ref{fig:netschematic}.

For the sake of brevity, our story omits an important aspect: Alice did not specify a perfect set of security goals at the first attempt. 
It took her some iteration to arrive at the specification shown in Figure~\ref{fig:runexsecinvars}. 
Fortunately, \topos{} has proven extremely helpful in this process. 
As described in the previous paragraph, once Alice instantiates a set of security invariant templates, \topos{} can compute a policy from them. 
This provides Alice with feedback about what she is specifying and what her specification actually means. 
In addition, \topos{} also allows Alice to define her own policy and \topos{} visualizes any flow which contradicts a specified security goal or highlights flows which Alice did not consider but which would be valid w.r.t.\ specified security goals.
Alice iterated this process several times and refined her security goals until she was certain that the specification carries the intended meaning. 

Due to this motivating feedback-driven process, Alice has now documented a clear and formal specification of her security requirements. 
The specification is modular and split into four invariants. 
Consequently, in the future, it is easy to add new security requirements or verify whether a proposed change violates the existing invariants.

\subsection{Policy Construction with \topos{}}
Given the specification of the security goals (Figure~\ref{fig:runexsecinvars}), \topos{} computes the security policy shown in Figure~\ref{fig:secpol}. 
Alice is happy with the policy, but she makes one small change. 
While the web frontend must be accessible from the Internet, there is currently no need that the web frontend also establishes connections to the Internet by itself. 
Alice modifies the policy as shown in Figure~\ref{fig:secpolrefined} and \topos{} verifies that her new policy complies with the security goals specified earlier. 

\begin{figure*}[htb]
	\centering
	\begin{minipage}{0.23\linewidth}\centering
		\resizebox{0.99\linewidth}{!}{%
	  	\begin{Large}
	    \begin{tikzpicture}
		   	\node[align=center,text width=8.5em,cloud, draw,cloud puffs=10,cloud puff arc=120, aspect=2, inner sep=-3em,outer sep=0] (a) at (3,0) { INET };
		   	\node (c) at (3,-2) { WebApp };
		   	\node (d) at (3,-4) { DB };
		   	\node (e) at (0,-3) { Log };
		   	\node (b) at (0,-1) { WebFrnt };
		   	
		   	\draw[myptr] (a) to[out=330,in=310,looseness=3] (a);
		   	\draw[myptr] (a) to (b);
		   	\draw[myptr] (b) to (a); %
		   	\draw[myptr] (b) to[loop above] (b);
		   	\draw[myptr] (b) to (c);
		   	\draw[myptr] (b) to (e);
		   	\draw[myptr] (c) to (a);
		   	\draw[myptr] (c) to (b);
		   	\draw[myptr] (c) to[loop right] (c);
		   	\draw[myptr] (c) to (d);
		   	\draw[myptr] (c) to (e);
		   	\draw[myptr] (d) to (c);
		   	\draw[myptr] (d) to[loop right] (d);
		   	\draw[myptr] (d) to (e);
		   	\draw[myptr] (e) to[loop below] (e);

	    \end{tikzpicture}
	    \end{Large}}
  		\vskip-5pt
		\caption[Security Policy computed]{Security Policy\newline{}\mbox{(computed by \topos{})}} %
		\label{fig:secpol}
	\end{minipage}
	\hspace*{\fill}
	\begin{minipage}{0.23\textwidth}\centering
		\resizebox{0.99\linewidth}{!}{%
	  	\begin{Large}
	    \begin{tikzpicture}
		   	\node[align=center,text width=8.5em,cloud, draw,cloud puffs=10,cloud puff arc=120, aspect=2, inner sep=-3em,outer sep=0] (a) at (3,0) { INET };
		   	\node (c) at (3,-2) { WebApp };
		   	\node (d) at (3,-4) { DB };
		   	\node (e) at (0,-3) { Log };
		   	\node (b) at (0,-1) { WebFrnt };
		   	
		   	\draw[myptr] (a) to[out=330,in=310,looseness=3] (a);
		   	\draw[myptr] (a) to (b);
		   	\draw[myptr] (b) to[loop above] (b);
		   	\draw[myptr] (b) to (c);
		   	\draw[myptr] (b) to (e);
		   	\draw[myptr] (c) to (a);
		   	\draw[myptr] (c) to (b);
		   	\draw[myptr] (c) to[loop right] (c);
		   	\draw[myptr] (c) to (d);
		   	\draw[myptr] (c) to (e);
		   	\draw[myptr] (d) to (c);
		   	\draw[myptr] (d) to[loop right] (d);
		   	\draw[myptr] (d) to (e);
		   	\draw[myptr] (e) to[loop below] (e);

	    \end{tikzpicture}
	    \end{Large}}
  		\vskip-5pt
		\caption[Security Policy refined]{Security Policy\newline{}\mbox{(manually refined)}}
		\label{fig:secpolrefined}
	\end{minipage}
	\hspace*{\fill}
	\begin{minipage}{0.23\linewidth}\centering
		\resizebox{0.99\linewidth}{!}{%
	  	\begin{Large}
	    \begin{tikzpicture}
		   	\node[align=center,text width=8.5em,cloud, draw,cloud puffs=10,cloud puff arc=120, aspect=2, inner sep=-3em,outer sep=0] (a) at (3,0) { INET };
		   	\node (c) at (3,-2) { WebApp };
		   	\node (d) at (3,-4) { DB };
		   	\node (e) at (0,-3) { Log };
		   	\node (b) at (0,-1) { WebFrnt };
		   	
		   	\draw[myptr] (a) to[out=330,in=310,looseness=3] (a);
		   	\draw[myptr] (a) to (b);
		   	\draw[myptr] (b) to[loop above] (b);
		   	\draw[myptr] (b) to (c);
		   	\draw[myptr] (b) to (e);
		   	\draw[myptr] (c) to (a);
		   	\draw[myptr] (c) to (b);
		   	\draw[myptr] (c) to[loop right] (c);
		   	\draw[myptr] (c) to (d);
		   	\draw[myptr] (c) to (e);
		   	\draw[myptr] (d) to (c);
		   	\draw[myptr] (d) to[loop right] (d);
		   	\draw[myptr] (d) to (e);
		   	\draw[myptr] (e) to[loop below] (e);
		   	
		   	\draw[myptr,dashed,TUMOrange] (a) to[bend left=15] (c);
		   	\draw[myptr,dashed,TUMOrange] ($(b) + (3ex,1.5ex)$) to[bend left=15] (a);
	    \end{tikzpicture}
	    \end{Large}}
  		\vskip-5pt
		\caption[Stateful Policy]{Stateful Policy\newline{}\mbox{(computed by \topos{})}}
		\label{fig:statefulpol}
	\end{minipage}
	\hspace*{\fill}
	\begin{minipage}{0.20\linewidth}\centering\centering
		\resizebox{\linewidth}{!}{
	  	\begin{LARGE}%
		\begin{tikzpicture}
		\node[align=center,text width=15.5em, cloud, draw,cloud puffs=10,cloud puff arc=110, aspect=2, inner sep=-3.5em,outer sep=0] (a) at (5,1) { $\{0.0.0.0 .. 9.255.255.255\} \cup \{11.0.0.0 .. 255.255.255.255\}$ };
		\node (b) at (5,-3) { $\{10.0.0.4\}$ };
		\node (c) at (5,-6) { $\{10.0.0.3\}$ }; %
		\node (d) at (0,-5) { $\{10.0.0.2\}$ };
		\node (e) at (0,-1) { $\{10.0.0.1\}$ };
		\node (f) at (4,-7.2) { $\{10.0.0.0\} \cup \{10.0.0.5 .. 10.255.255.255\}$ };
		
		\draw[myptr] (a) to[out=330,in=310,looseness=3] (a);
		\draw[myptr] (a) to (b);
		\draw[myptr] (a) to[bend left] (c);
		\draw[myptr] (a) to (d);
		\draw[myptr] (a) to (e);
		\draw[myptr] (a) to[bend right] ($(f.north)+(-4ex,0)$);
		\draw[myptr] (b) to (a);
		\draw[myptr] (b) to[loop right] (b);
		\draw[myptr] (b) to (c);
		\draw[myptr] (b) to (d);
		\draw[myptr] (b) to (e);
		\draw[myptr] (b) to[bend right] ($(f.north)+(-4ex,0)$);
		\draw[myptr] (c) to (b);
		\draw[myptr] (c) to[loop right] (c);
		\draw[myptr] (c) to (d);
		\draw[myptr] (d) to[loop below] (d);
		\draw[myptr] (e) to (b);
		\draw[myptr] (e) to (d);
		\draw[myptr] (e) to[loop above] (e);
		\end{tikzpicture}%
		\end{LARGE}%
		}
		\vskip-1ex
		\caption{Firewall Overview\newline{}\mbox{(reconstructed by \fffuu{})}}\label{fig:mansdn:docker:servicematrix}
	\end{minipage}
\end{figure*}

A careful reader may already notice the subtle semantics we gave to the arrows in Figure~\ref{fig:secpolrefined}: 
For example, the arrow from the Internet to the web frontend means that the Internet may set up connections to the web frontend, but not vice versa.  
A connection usually implies that---once it is established---packets may flow in both directions. 
But the policy does currently not permit packets to flow from the web frontend to the Internet. 
Therefore, using the specification of the security goals again, \topos{} can convert a security policy into a stateful policy~\cite{diekmann2014EPTCS}. 
The result is shown in Figure~\ref{fig:statefulpol}. 
The orange dashed arrows represent connections which may be stateful, \ie once established by the corresponding entity (solid black arrow), packets may travel in both directions (solid black + dashed orange arrow). 

To compute the stateful policy, \topos{} distinguishes between access control (invariants 1 and 4) and information flow (invariants 2 and 3). 
This distinction is inherently built into the invariant templates~\cite{diekmann2014forte} and Alice does not need to configure anything for this step. 
In this scenario, the two stateful flows can be justified as they do not introduce access control violations since the initiator needs to establish a connection before a bidirectional flow is allowed. 
However, it can be seen that all flows which send data to the logging server are not allowed to be stateful. 
This is because of information flow security: a bidirectional channel here would allow the confidential logging information to leak. 
\section{Deploying to a Docker Host}
\label{sec:deploytodocker}
Now, Alice wants to enforce the stateful policy of Figure~\ref{fig:statefulpol} on her Docker host. 
Docker uses the iptables firewall internally. 
The first question Alice raises is whether she wants to operate custom firewall rules or whether Docker can enforce her policy out of the box. 

Since Docker version 1.10, it is possible to create custom internal networks~\cite{docker2016blog110}. 
The \texttt{{-}{-}internal} flag protects a network from accesses from the outside. 
However, by default, all containers in an internal network can reach each other. 
It is possible to configure the network such that containers cannot access each other. 
Unfortunately, a bug in Alice's version of Docker allowed containers to communicate~\cite{github2016dockerinternalicc}. 
In addition, there was no possibility to enable fine-grained access control between the containers in a network. 
A \texttt{{-}{-}link} option exists to connect two containers within a custom network, but it merely sets environment variables, it does not influence the actual IP connectivity. 
The Docker design philosophy is to decouple the application developer from networking details~\cite{docker2016blognetphilosophy}. 
Only a coarse-grained network abstraction in terms of different networks is exposed to the application developer. 
The network IT team ---\ie Alice--- should manage the network. 
Since one compromised container in an internal network can attack all other containers in its network, Alice desires further network-level access control than the coarse grained separation provided by Docker. %

The web application as a whole should be isolated from other containers on the host. 
Therefore, Alice creates a new Docker network. 
She disables inter-container communication (icc) for this network. 
Due to the lack of fine-grained access control, Alice decides to install additional custom firewall rules on her host. 
In addition, the fact that the fix to the bug mentioned above was initially not considered a security issue by the Docker developers confirms Alice's choice of running her own firewall. 
As it turns out, many administrators are looking for means to fine-tune their Docker firewall because of certain shortcomings in Docker~\cite{github2016dockerfwbypass}. 
Alice chooses to enhance the default Docker-generated rules with rules for fine-grained access control generated by \topos{}. 
She merges both rulesets manually. 
Figure~\ref{fig:dockerfirewall:tuned} shows the final ruleset of her Docker host. 
Everything which does not mention \texttt{MYNET} is part of the default ruleset of docker. 
Alice added the custom chain \texttt{MYNET} and a jump to it. 
The contents of the chain \texttt{MYNET}, \ie the actual fine-grained access control rules, are generated by \topos{}. 
For brevity, the name of the bridging interface created by Docker for Alice's new network is \texttt{\dockerbr{}}. 

\begin{figure}[!htb]
	\centering
	\fbox{%
		\hspace*{1ex}
		\resizebox{0.9\linewidth}{!}{%
		\begin{minipage}{\widthof{\verb~-A MYNET -i \dockerbr{} -s 10.0.0.1 -o \dockerbr{} -d 10.0.0.1 -j ACCEPT   ~}}%
			\input{notreallyallinone_sdnnfv_dockerfirewall_tuned_fig.tex}
		\end{minipage}%
		}
		\hspace*{1ex}			
	}\caption{Docker \texttt{iptables} firewall (only \texttt{filter} table shown).}\label{fig:dockerfirewall:tuned}
\end{figure}

As a side note, Alice is well aware that Docker itself does not provide security out of the box~\cite{dockersecurity,nccdocker2016security}. 
The security provided by containerization heavily depends on the specific Docker and container setup. 
For example, if a container has the privilege to access raw sockets, the container can spoof arbitrary IP addresses or perform ARP spoofing~\cite{bui15arxivsockersec}. 
This spoofing cannot be blocked by traditional firewall rules since all containers in Alice's network are connected to the same interface \texttt{\dockerbr{}}. 
Hence, in a generic case, Docker networking does not provide authenticity. 
Consequently, since a malicious, privileged container can perform ARP spoofing, no guarantees can be given that the connectivity structure enforced by Alice's firewall actually corresponds to Figure~\ref{fig:statefulpol}. 
Alice solves this problem by teaching her application developers about Linux capabilities and rejecting containers which require any additional capabilities or are otherwise `naked'~\cite{jessfraznakedcontainer}. 

\section{Operations}
\label{sec:operations}
The design and development of the firewall policy are only the initial phase of the app's life cycle. 
After the firewall has been deployed, additional and changing requirements necessitate updates to the previously designed ruleset. 
In contrast to network design and tool development, operations is usually an unstructured ad-hoc discipline which has to deal with unexpected events and usually is not taught in any course of studies~\cite{goldfuss2016passing}. 
In this section, we follow Alice while she deals with unexpected events, ad-hoc requests, overlapping issues, and long-term improvements of her setup within limited time budget.

Alice used \topos{} to help her with \emph{static} configuration management. 
Consequently, Alice now needs tools to help her in \emph{dynamic} contexts. 
Alice installs a cronjob which runs \texttt{iptables-save} regularly and looks for changes. 
If a change to the ruleset is discovered, \fffuu{} is run to compute an overview of the firewall policy currently enforced.  
The result is visualized with tikz and emailed to Alice. 
The result for the initial firewall rules is visualized in Figure~\ref{fig:mansdn:docker:servicematrix}. 

For an deeper look into operating principles and implementation of \fffuu{} we refer to the original publication \cite{diekmann2016networking}.

Since \fffuu{} operates on the raw iptables rules, it does not know the hostnames of the containers. 
Further comparing the designed policy in Figure~\ref{fig:statefulpol} with Figure~\ref{fig:mansdn:docker:servicematrix}, Alice realizes that her firewall policy is already non-optimal. 
The reason is that the default Docker-generated rules are interacting with her rules. 
One oddity in \fffuu{}'s visualization is the IP range at the bottom which corresponds to all unused IPs in Alice's new network 10.0.0.0/24. 
Those IPs are potentially accessible from the Internet and the \mbox{$\mvar{WebApp}$}. 
Since no containers are assigned to these IP addresses, Alice decides that this is acceptable for now. 
For her, a more concerning observation is that the Internet can access almost all internal containers. 
But Alice knows that she does not export internal services to the Internet in her Docker configuration, which is not visible in the firewall setup. 
Due to time pressure, Alice decides that the current firewall rules are good enough and she will add a second line of defense later. 
This will also protect against attacks from containers running on the same host but in a different Docker network. 
Alice runs one last check with \fffuu{}, verifying that $\mvar{INET} \rightarrow \mvar{WebFrnt}$ (10.0.0.1) is the only stateful flow; as discussed, $\mvar{WebApp} \rightarrow \mvar{INET}$ is already allowed bidirectionally. 
Due to time constraints, Alice decides not to ask \fffuu{} about stateful flows furthermore.

On Friday afternoon, Alice receives a call from a friend at \url{heise.de} (193.99.144.80), who is complaining that one of her containers is pinging his webserver excessively. 
Alice knows that the web backend tests connectivity from time to time by sending one echo request to Heise. 
She decides to postpone investigating the core of the problem and installs a short-term mitigation by just rate limiting all connections to Heise. 
We print changes to the ruleset in unified \texttt{diff} format. 
Alice installs the following rules. 

\smallskip

\begin{minipage}{.95\linewidth}
	\footnotesize
	\begin{Verbatim}[commandchars=\\\{\},codes={\catcode`$=3\catcode`^=7}]
 :DOCKER-ISOLATION - [0:0]
 :MYNET - [0:0]
 -A FORWARD -j DOCKER-ISOLATION
\diffadd{+-A FORWARD -d 193.99.144.80 -m recent} $\hfill\hookleftarrow$
     \diffadd{--set --name rateheise --rsource}
\diffadd{+-A FORWARD -d 193.99.144.80 -m recent} $\hfill\hookleftarrow$
     \diffadd{--update --seconds 60 --hitcount 3} $\hfill\hookleftarrow$
     \diffadd{--name rateheise --rsource -j DROP}
 -A FORWARD -j MYNET
 -A FORWARD -o \dockerbr{} -j DOCKER
 -A FORWARD -o \dockerbr{} -m conntrack $\hfill\hookleftarrow$
    --ctstate RELATED,ESTABLISHED -j ACCEPT
\end{Verbatim}
\end{minipage}
\medskip

While this change to the ruleset is identified by her cronjob, \fffuu{} confirms that the overall access control structure of the firewall has not changed: It still corresponds to Figure~\ref{fig:mansdn:docker:servicematrix}. 

While Alice is updating the ruleset, she remembers from Figure~\ref{fig:mansdn:docker:servicematrix} that the Internet still has too many direct access rights to her internal containers. 
While working on the ruleset, Alice decides to fix this issue right away mow. %

\smallskip

\begin{minipage}{.95\linewidth}
	\footnotesize
	\begin{Verbatim}[commandchars=\\\{\},codes={\catcode`$=3\catcode`^=7}]
 -A MYNET -i \dockerbr{} -s 10.0.0.4 -o \dockerbr{} -d 10.0.0.3 $\hfill\hookleftarrow$
     -j ACCEPT
 -A MYNET -i \dockerbr{} -s 10.0.0.4 -o \dockerbr{} -d 10.0.0.2 $\hfill\hookleftarrow$
     -j ACCEPT
 -A MYNET -i \dockerbr{} -s 10.0.0.4 -o \dockerbr{} -d 10.0.0.4 $\hfill\hookleftarrow$
     -j ACCEPT
\diffdel{--A MYNET -i \dockerbr{} -s 10.0.0.4 ! -o \dockerbr{} -j ACCEPT}
\diffdel{--A MYNET ! -i \dockerbr{} -o \dockerbr{} -d 10.0.0.1 -j ACCEPT}
\diffadd{+-A MYNET -i \dockerbr{} -s 10.0.0.4} $\hfill\hookleftarrow$
     \diffadd{! -o \dockerbr{} ! -d 10.0.0.0/8 -j ACCEPT}
\diffadd{+-A MYNET ! -i \dockerbr{} ! -s 10.0.0.0/8} $\hfill\hookleftarrow$
     \diffadd{-o \dockerbr{} -d 10.0.0.1 -j ACCEPT}
 -A MYNET -i \dockerbr{} -j DROP
\diffadd{+-A MYNET -o \dockerbr{} -j DROP}
\diffadd{+-A MYNET -s 10.0.0.0/8 -j DROP}
\diffadd{+-A MYNET -d 10.0.0.0/8 -j DROP}
 COMMIT
	\end{Verbatim}
\end{minipage}

\medskip

After a few seconds, she receives the new firewall overview generated by \fffuu{}, shown in Figure~\ref{fig:dynamicdocker:fffuu:dockermynet3}. 
Comparing this figure %
  to Figure~\ref{fig:mansdn:docker:servicematrix}, it can be seen that the Internet is now appropriately constrained. 
Alice also happily realizes that her firewall ruleset now corresponds to the original design of Figure~\ref{fig:statefulpol}.

\begin{figure}[htb]
	\centering
	\begin{minipage}[t]{0.45\linewidth}\centering
		\resizebox{\linewidth}{!}{\Large
		\begin{tikzpicture}
		\node[align=center,text width=15.5em, cloud, draw,cloud puffs=10,cloud puff arc=120, aspect=2, inner sep=-3em,outer sep=0] (a) at (5,1) { $\{0.0.0.0 .. 9.255.255.255\} \cup \{11.0.0.0 .. 255.255.255.255\}$ };
		\node (b) at (5,-3) { $\{10.0.0.4\}$ };
		\node (c) at (5,-6) { $\{10.0.0.3\}$ }; %
		\node (d) at (0,-5) { $\{10.0.0.2\}$ };
		\node (e) at (0,-1) { $\{10.0.0.1\}$ };
		\node (f) at (4,-8) { $\{10.0.0.0\} \cup \{10.0.0.5 .. 10.255.255.255\}$ };
		
		\draw[myptr] (a) to[out=330,in=310,looseness=3] (a);
		\draw[myptr] (a) to (e);
		\draw[myptr] (b) to (a);
		\draw[myptr] (b) to[loop right] (b);
		\draw[myptr] (b) to (c);
		\draw[myptr] (b) to (d);
		\draw[myptr] (b) to (e);
		\draw[myptr] (c) to (b);
		\draw[myptr] (c) to[loop right] (c);
		\draw[myptr] (c) to (d);
		\draw[myptr] (d) to[loop below] (d);
		\draw[myptr] (e) to (b);
		\draw[myptr] (e) to (d);
		\draw[myptr] (e) to[loop above] (e);
		\end{tikzpicture}%
		}
	\caption{Overview computed by \fffuu{} after call from Heise}
	\label{fig:dynamicdocker:fffuu:dockermynet3}
	\end{minipage}%
	\hspace*{\fill}%
	\begin{minipage}[t]{0.45\linewidth}\centering
	   \resizebox{\linewidth}{!}{\Large
		\begin{tikzpicture}
		\node[align=center,text width=15.5em,cloud, draw,cloud puffs=10,cloud puff arc=120, aspect=2, inner sep=-3em,outer sep=0] (a) at (5,1) { $\{0.0.0.0 .. 9.255.255.255\} \cup \{11.0.0.0 .. 255.255.255.255\}$ };
		\node (b) at (5,-3) { $\{10.0.0.4\}$ };
		\node (c) at (5,-6) { $\{10.0.0.3\}$ }; %
		\node (d) at (0,-5) { $\{10.0.0.2\}$ };
		\node (e) at (0,-1) { $\{10.0.0.1\}$ };
		\node (f) at (4,-8) { $\{10.0.0.0\} \cup \{10.0.0.5 .. 10.255.255.255\}$ };
	
		\draw[myptr] (a) to[out=330,in=310,looseness=3] (a);
		\draw[myptr] (a) to (e);
		\draw[myptr] (b) to (a);
		\draw[myptr] (b) to[loop right] (b);
		\draw[myptr] (b) to (c);
		\draw[myptr] (b) to (d);
		\draw[myptr] (b) to (e);
		\draw[myptr] (c) to (b);
		\draw[myptr] (c) to[loop right] (c);
		\draw[myptr] (c) to (d);
		\draw[myptr] (d) to[loop below] (d);
		\draw[myptr] (d) to (e);
		\draw[myptr] (e) to (b);
		\draw[myptr] (e) to (d);
		\draw[myptr] (e) to[loop above] (e);
	\end{tikzpicture}%
	}
	\caption{Overview computed by \fffuu{} after WebDev call (HTTP)} %
	\label{fig:dynamicdocker:fffuu:dockermynet4}
	\end{minipage}%
\end{figure}

The next week, Alice receives a call from her web developer. 
He requests ssh access to all containers. 
In addition, he requests that the log server (10.0.0.2) may access a status page of the web frontend (10.0.0.1) over HTTP. 
Both permissions should only be granted temporarily for debugging purposes. 
Alice sets up the appropriate firewall rules. 

\smallskip

\begin{minipage}{.95\linewidth}
	\footnotesize
	\begin{Verbatim}[commandchars=\\\{\},codes={\catcode`$=3\catcode`^=7}]
 -A FORWARD -j DOCKER-ISOLATION
 -A FORWARD -d 193.99.144.80 -m recent $\hfill\hookleftarrow$
     --set --name rateheise --rsource
 -A FORWARD -d 193.99.144.80 -m recent $\hfill\hookleftarrow$
     --update --seconds 60 --hitcount 3 $\hfill\hookleftarrow$
     --name rateheise --rsource -j DROP
\diffadd{+-A FORWARD -m state }$\hfill\hookleftarrow$
     \diffadd{--state ESTABLISHED,RELATED -j ACCEPT}
\diffadd{+-A FORWARD -p tcp --dport 22 -j ACCEPT}
\diffadd{+-A FORWARD -s 10.0.0.2 -d 10.0.0.1 -p tcp} $\hfill\hookleftarrow$
     \diffadd{--dport 80 -j ACCEPT}
 -A FORWARD -j MYNET
 -A FORWARD -o \dockerbr{} -j DOCKER
 -A FORWARD -o \dockerbr{} -m conntrack $\hfill\hookleftarrow$
     --ctstate RELATED,ESTABLISHED -j ACCEPT
	\end{Verbatim}
\end{minipage}

\medskip

While the Docker container connectivity works as intended, \fffuu{} now computes two interesting access control overviews.\footnote{By default, \fffuu{} only computes the overview for a fixed service, by default ssh and HTTP. } 
First, it visualizes that there are no longer any restrictions for ssh, \ie \fffuu{} only shows one node. 
This single node comprises the complete IPv4 address space and may access itself, \ie unconstrained ssh access. %
Second, \fffuu{} presents a new overview of the HTTP connectivity, shown in Figure~\ref{fig:dynamicdocker:fffuu:dockermynet4}. 
The only difference is that the log server may now access the web frontend. 
The two overviews underline that Alice implemented her web developer's request correctly.

While the policy overview computed by \fffuu{} is still comprehensible for Alice due to the pooling of equivalent IP addresses, the raw firewall ruleset, now comprising 37 rules, is slowly becoming a mess\footnote{In a study of 15 real world iptables rulesets with up to 4946 rules, we observed at most 13 groups of IP addresses for the SSH or HTTP \cite{diekmann2016networking}. For a specific group, the behaviour can be determined by the interaction with the remaining groups}. 
The ruleset partly contains unused artifacts installed by Docker and Alice's hot fixes are cluttered all over it. 
While Alice is poring over about how she could clean up the rules, she receives an emergency call. 
Her manager tells her that the webservice was mentioned on reddit and that the web frontend cannot cope with the increased load. 
Alice is spawning an additional frontend container with IP address 10.0.0.42. 

Alice's firewall adheres to best practices and implements whitelisting. 
Consequently, the new container does not have any connectivity. 
Alice now is in the urgent situation to get the firewall rules set up which permit connectivity for the second frontend instance. 
Just permitting everything is not an acceptable option for security-aware Alice.

\begin{figure}[htb]
	\centering
	\begin{minipage}[b]{0.49\linewidth}\centering
		\centering
		\includegraphics[trim=0cm 1cm 0cm 0cm,clip=true,width=\textwidth]{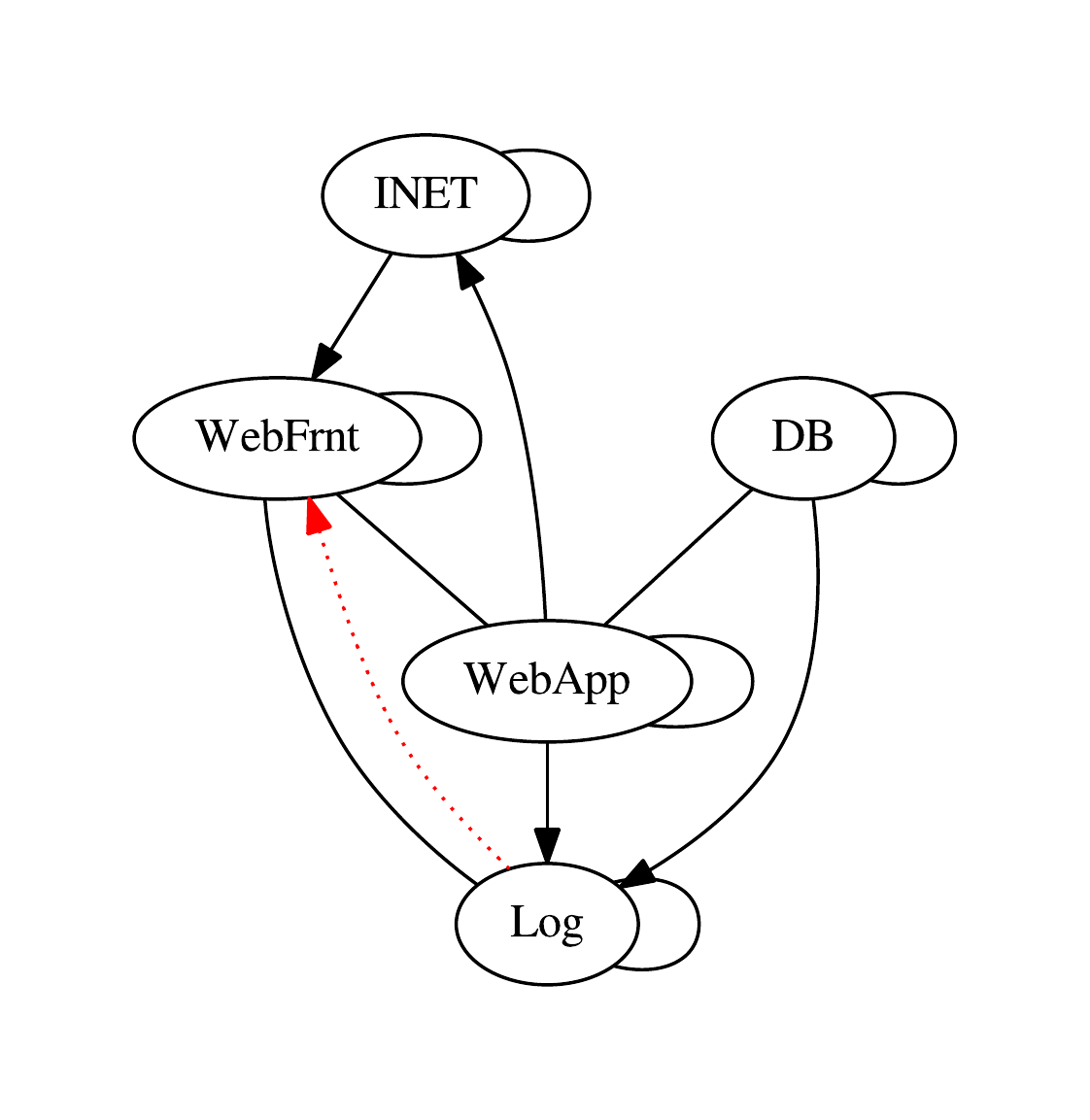}
		\vskip-1ex
		\caption{Uncovered violation\newline{}\mbox{(screenshot of \topos{})}}
		\label{fig:dynamicdocker:topos:dockermynet4}
	\end{minipage}%
	\hspace*{\fill}%
	\begin{minipage}[b]{0.49\linewidth}\centering
		\centering
		\includegraphics[trim=0cm 1cm 0cm 0cm,clip=true,width=\textwidth]{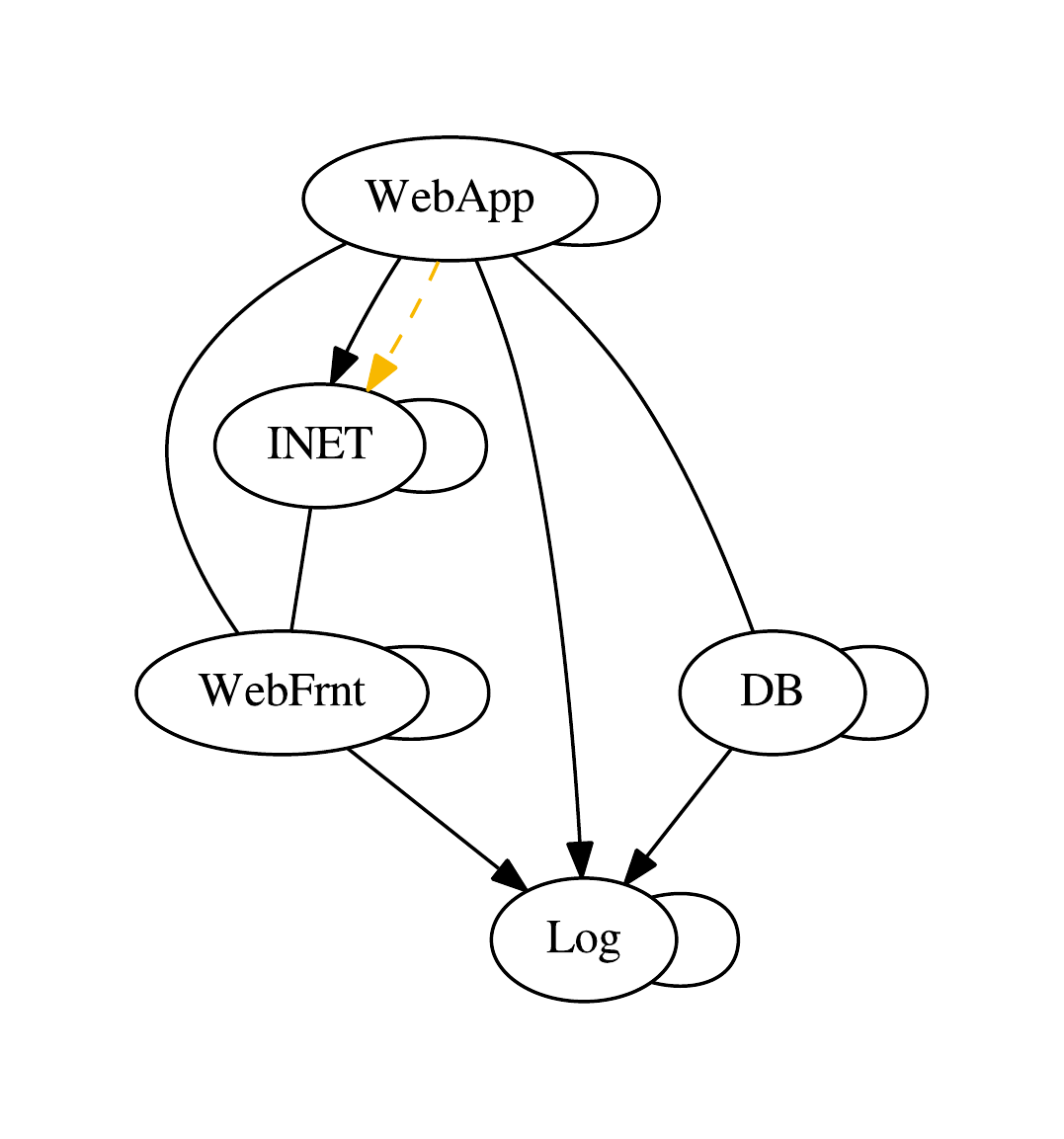}
		\vskip-1ex
		\caption{Stateful Policy\newline{}\mbox{(screenshot of \topos{})}}
		\label{fig:dynamicdocker:topos:dockermynet4.stateful}
	\end{minipage}%
\end{figure}

Fortunately, Alice remembers that she has previously specified the security requirements with \topos{} (Figure~\ref{fig:runexsecinvars}). 
First, Alice loads the policy overview for HTTP computed by \fffuu{} (Figure~\ref{fig:dynamicdocker:fffuu:dockermynet4}) into \topos{}. 
She immediately gets a visualization (Figure~\ref{fig:dynamicdocker:topos:dockermynet4}) which tells her that her policy has already diverged from the security requirements: The dotted red arrow, visualized by \topos{}, reveals a flow which is violating the security requirements. 
It corresponds to the flow installed upon the request of her web developer previously.  
Based on the security requirements, Alice should probably have rejected the request from her web developer in the very beginning. 
But this is neither the time to cast the blame nor to discuss security requirements. 
Given the time pressure and the many odd-looking Docker-generated rules in her ruleset, Alice decides that her current ruleset is not salvageable. 
Alice is not the first person to notice that the Docker daemon may make surprising changes to an iptables configuration~\cite{docker2014blogufw,github2016dockerfwbypass}. 
She decides to take over complete control over the ruleset. 
Therefore, Alice prohibits Docker from making any changes to the firewall by setting \texttt{{-}{-}iptables=false}.

\begin{figure}[htbp]
	\begin{minipage}{\linewidth}
	\scriptsize
\begin{Verbatim}[commandchars=\\\{\},codes={\catcode`$=3\catcode`^=7}]
-A FORWARD -i \dockerbr{} -s $\mathit{\$WebFrnt\ippostfix{}}$ -o \dockerbr{} -d $\mathit{\$WebFrnt\ippostfix{}}$ -j ACCEPT 
-A FORWARD -i \dockerbr{} -s $\mathit{\$WebFrnt\ippostfix{}}$ -o \dockerbr{} -d $\mathit{\$Log\ippostfix{}}$ -j ACCEPT 
-A FORWARD -i \dockerbr{} -s $\mathit{\$WebFrnt\ippostfix{}}$ -o \dockerbr{} -d $\mathit{\$WebApp\ippostfix{}}$ -j ACCEPT 
-A FORWARD -i \dockerbr{} -s $\mathit{\$WebFrnt\ippostfix{}}$ -o $\mathit{\$INET\_iface}$ -d $\mathit{\$INET\ippostfix{}}$ $\hfill\hookleftarrow$
    -j ACCEPT 
-A FORWARD -i \dockerbr{} -s $\mathit{\$DB\ippostfix{}}$ -o \dockerbr{} -d $\mathit{\$DB\ippostfix{}}$ -j ACCEPT 
-A FORWARD -i \dockerbr{} -s $\mathit{\$DB\ippostfix{}}$ -o \dockerbr{} -d $\mathit{\$Log\ippostfix{}}$ -j ACCEPT 
-A FORWARD -i \dockerbr{} -s $\mathit{\$DB\ippostfix{}}$ -o \dockerbr{} -d $\mathit{\$WebApp\ippostfix{}}$ -j ACCEPT 
-A FORWARD -i \dockerbr{} -s $\mathit{\$Log\ippostfix{}}$ -o \dockerbr{} -d $\mathit{\$Log\ippostfix{}}$ -j ACCEPT 
-A FORWARD -i \dockerbr{} -s $\mathit{\$WebApp\ippostfix{}}$ -o \dockerbr{} -d $\mathit{\$WebFrnt\ippostfix{}}$ -j ACCEPT 
-A FORWARD -i \dockerbr{} -s $\mathit{\$WebApp\ippostfix{}}$ -o \dockerbr{} -d $\mathit{\$DB\ippostfix{}}$ -j ACCEPT 
-A FORWARD -i \dockerbr{} -s $\mathit{\$WebApp\ippostfix{}}$ -o \dockerbr{} -d $\mathit{\$Log\ippostfix{}}$ -j ACCEPT 
-A FORWARD -i \dockerbr{} -s $\mathit{\$WebApp\ippostfix{}}$ -o \dockerbr{} -d $\mathit{\$WebApp\ippostfix{}}$ -j ACCEPT 
-A FORWARD -i \dockerbr{} -s $\mathit{\$WebApp\ippostfix{}}$ -o $\mathit{\$INET\_iface}$ -d $\mathit{\$INET\ippostfix{}}$ $\hfill\hookleftarrow$
    -j ACCEPT 
-A FORWARD -i $\mathit{\$INET\_iface}$ -s $\mathit{\$INET\ippostfix{}}$ -o \dockerbr{} -d $\mathit{\$WebFrnt\ippostfix{}}$ $\hfill\hookleftarrow$
    -j ACCEPT 
-A FORWARD -i $\mathit{\$INET\_iface}$ -s $\mathit{\$INET\ippostfix{}}$ $\hfill\hookleftarrow$ 
    -o $\mathit{\$INET\_iface}$ -d $\mathit{\$INET\ippostfix{}}$ -j ACCEPT 
-I FORWARD -m state --state ESTABLISHED -i $\mathit{\$INET\_iface}$ $\hfill\hookleftarrow$
    -s $\mathit{\$INET\ippostfix{}}$ -o \dockerbr{} -d $\mathit{\$WebApp\ippostfix{}}$ -j ACCEPT 
\end{Verbatim}
\end{minipage}%
\caption{Fresh ruleset generated by \topos{}, considering only the requirements}
\label{fig:dynamicdocker:topos:dockermynet.topos4}
\end{figure}

To start over, Alice asks \topos{} to compute a completely new ruleset for her, based only on the requirements specified before. 
\topos{} computes the stateful policy shown in Figure~\ref{fig:dynamicdocker:topos:dockermynet4.stateful}; the dashed orange flow indicates a connection with stateful semantics. 
The result is serialized to the firewall rules shown in Figure~\ref{fig:dynamicdocker:topos:dockermynet.topos4}. 
All containers are attached to the same Docker bridge. %
Alice only needs to fill in the IP addresses of the machines and the Internet-facing interface. 
Alice sticks to the IP addresses of her containers as before, except for the web frontend which now has two active containers running. 
She sets $\mathit{\$WebFrnt\ippostfix{}} = \textnormal{10.0.0.1,10.0.0.42}$. 
This syntax is supported by iptables and iptables will automatically expand this syntax to several rules upon loading. 
To specify the interface and IP range of the Internet, Alice defines that the Internet is `everything except the Docker subnet'. 
Therefore, Alice negates her Docker interface and internal Docker IP range. 
\begin{samepage}
For example, the second last rule becomes the following: 

\begin{footnotesize}
\begin{Verbatim}[commandchars=\\\{\},codes={\catcode`$=3\catcode`^=7}]
-A FORWARD ! -i \dockerbr{} ! -s 10.0.0.0/8 $\hfill\hookleftarrow$
    ! -o \dockerbr{} ! -d 10.0.0.0/8 -j ACCEPT
\end{Verbatim}
\end{footnotesize}%
\end{samepage}

Yet, for some rules, the iptables command refuses to load her rules and complains that negation is not allowed with multiple source or destination IP addresses. 
For example in line four, iptables prohibits the use of \verb~! -d 10.0.0.0/8~ in combination with the two source addresses \verb~-s 10.0.0.1,10.0.0.42~ specified for $\mathit{\$WebFrnt\ippostfix{}}$. 
To work around this iptables limitation, Alice uses the \verb~iprange~ module to declare the IP range of the Internet. 
For example, the fourth rule now becomes

\begin{footnotesize}
\begin{Verbatim}[commandchars=\\\{\},codes={\catcode`$=3\catcode`^=7}]
-A FORWARD -i \dockerbr{} -s 10.0.0.1,10.0.0.42 ! -o \dockerbr{} $\hfill\hookleftarrow$
    -m iprange ! --dst-range $\hfill\hookleftarrow$
    10.0.0.0-10.255.255.255 -j ACCEPT
\end{Verbatim}
\end{footnotesize}

\begin{figure}[htb]
	\centering
	\begin{minipage}[t]{0.45\linewidth}\centering
		\resizebox{\linewidth}{!}{\Large
		\begin{tikzpicture}
		\node[align=center,text width=15.5em,cloud, draw,cloud puffs=10,cloud puff arc=120, aspect=2, inner sep=-3em,outer sep=0] (a) at (5,1) { $\{0.0.0.0 .. 9.255.255.255\} \cup \{11.0.0.0 .. 255.255.255.255\}$ };
		\node (c) at (5,-3) { $\{10.0.0.4\}$ };
		\node (d) at (5,-6) { $\{10.0.0.3\}$ }; %
		\node (e) at (0,-5) { $\{10.0.0.2\}$ };
		\node (b) at (0,-1) { $\{10.0.0.1,10.0.0.42\}$ };
		\node[align=center,text width=15.5em] (f) at (3,-8) { $\{10.0.0.0\} \cup \{10.0.0.5 .. 10.0.0.41\} \cup \{10.0.0.43 .. 10.255.255.255\} $ };
		
		\draw[myptr] (a) to[out=330,in=310,looseness=3] (a);
		\draw[myptr] (a) to (b);
		\draw[myptr] (b) to (a);
		\draw[myptr] (b) to[loop above] (b);
		\draw[myptr] (b) to (c);
		\draw[myptr] (b) to (e);
		\draw[myptr] (c) to (a);
		\draw[myptr] (c) to (b);
		\draw[myptr] (c) to[loop right] (c);
		\draw[myptr] (c) to (d);
		\draw[myptr] (c) to (e);
		\draw[myptr] (d) to (c);
		\draw[myptr] (d) to[loop right] (d);
		\draw[myptr] (d) to (e);
		\draw[myptr] (e) to[loop below] (e);
		\end{tikzpicture}%
		}
	\caption{Overview of Figure~\ref{fig:dynamicdocker:topos:dockermynet.topos4} computed by \fffuu{}} %
	\label{fig:dynamicdocker:fffuu:dockermynet.topos4}
	\end{minipage}%
	\hspace*{\fill}%
	\begin{minipage}[t]{0.45\linewidth}\centering
	   \resizebox{\linewidth}{!}{\Large
	   \begin{tikzpicture}
	   	\node[align=center,text width=15.5em,cloud, draw,cloud puffs=10,cloud puff arc=120, aspect=2, inner sep=-3em,outer sep=0] (a) at (5,1) { $\{0.0.0.0 .. 9.255.255.255\} \cup \{11.0.0.0 .. 255.255.255.255\}$ };
	   	\node (c) at (5,-3) { $\{10.0.0.4\}$ };
	   	\node (d) at (5,-6) { $\{10.0.0.3\}$ }; %
	   	\node (e) at (0,-5) { $\{10.0.0.2\}$ };
	   	\node (b) at (0,-1) { $\{10.0.0.1,10.0.0.42\}$ };
	   	\node[align=center,text width=15.5em] (f) at (3,-8) { $\{10.0.0.0\} \cup \{10.0.0.5 .. 10.0.0.41\} \cup \{10.0.0.43 .. 10.255.255.255\} $ };
	   	
	   	\draw[myptr] (a) to[out=330,in=310,looseness=3] (a);
	   	\draw[myptr] (a) to (b);
	   	\draw[myptr] (b) to (a);
	   	\draw[myptr] (b) to[loop above] (b);
	   	\draw[myptr] (b) to (c);
	   	\draw[myptr] (b) to (e);
	   	\draw[myptr] (c) to (a);
	   	\draw[myptr] (c) to (b);
	   	\draw[myptr] (c) to[loop right] (c);
	   	\draw[myptr] (c) to (d);
	   	\draw[myptr] (c) to (e);
	   	\draw[myptr] (d) to (c);
	   	\draw[myptr] (d) to[loop right] (d);
	   	\draw[myptr] (d) to (e);
	   	\draw[myptr] (e) to[loop below] (e);
	   	
	   	\draw[myptr,dashed,TUMOrange] (a) to[bend left=15, shorten <=0.6em,shorten >=0.2em] (c);
	   	\end{tikzpicture} %
	   	}
	\caption{HTTP access control overview with state (by \fffuu{})}
	\label{fig:dynamicdocker:fffuu:dockermynet.topos4.1.state}
	\end{minipage}%
\end{figure}

This loads fine. 
Fortunately, \fffuu{} understands those matching modules. 
The new firewall overview is visualized in Figure~\ref{fig:dynamicdocker:fffuu:dockermynet.topos4}. 
It is remarkably similar to Figure~\ref{fig:dynamicdocker:fffuu:dockermynet3}, the last, old visualization when the ruleset was still in a good state. 
The main difference is that the web frontend is now represented by two machines and that it may establish connections to the Internet itself. 
This has been prohibited by the old manually-refined policy, but it does not contradict any security requirement. 
A final test confirms that the container connectivity works as expected and the two frontend instances can cope with the load.

Looking at her todo list, Alice decides to install some old rules again. 
This time, she designs a clean ruleset and handles all of her temporary rules in a chain she calls \texttt{CUSTOM}. 
After her custom chain, she hands over control to the \topos{}-generated rules. 
Alice still has not investigated why some container is excessively pinging 193.99.144.80, so she installs the rate limiting again. 
Alice is more careful about the other temporary rules. 
\topos{} has shown her that the log server must not communicate with the web frontend, so she is not enabling this rule. 
However, since her ssh server is securely configured, she does not see a problem with the ssh exception and enables it again. 
She installs the following rules:

\smallskip

\begin{minipage}{.95\linewidth}
	\footnotesize
	\begin{Verbatim}[commandchars=\\\{\},codes={\catcode`$=3\catcode`^=7}]
 :INPUT ACCEPT [0:0]
 :FORWARD DROP [0:0]
 :OUTPUT ACCEPT [0:0] +:CUSTOM - [0:0]
\diffadd{+-A FORWARD -j CUSTOM}
\diffadd{+-A CUSTOM -d 193.99.144.80 -m recent}$\hfill\hookleftarrow$
\diffadd{      --set --name rateheise --rsource}
\diffadd{+-A CUSTOM -d 193.99.144.80 -m recent}$\hfill\hookleftarrow$
\diffadd{     --update --seconds 60 --hitcount 3} $\hfill\hookleftarrow$
     \diffadd{--name rateheise --rsource -j DROP}
\diffadd{+-A CUSTOM -m state --state ESTABLISHED -j ACCEPT}
\diffadd{+-A CUSTOM -p tcp -m tcp --dport 22 -j ACCEPT}
 -A FORWARD -i \dockerbr{} -s 10.0.0.1,10.0.0.42  $\hfill\hookleftarrow$
     -o \dockerbr{} -d 10.0.0.1,10.0.0.42 -j ACCEPT 
 -A FORWARD -i \dockerbr{} -s 10.0.0.1,10.0.0.42  $\hfill\hookleftarrow$
     -o \dockerbr{} -d 10.0.0.2 -j ACCEPT 
 -A FORWARD -i \dockerbr{} -s 10.0.0.1,10.0.0.42  $\hfill\hookleftarrow$
     -o \dockerbr{} -d 10.0.0.4 -j ACCEPT
	\end{Verbatim}
\end{minipage}

\medskip

Alice knows that it is a good practice to have a rule which allows all packets belonging to an established connection~\cite{iptablesperfectruleset}. 
She definitely needs an \verb~ESTABLISHED~ rule to make ssh work so she just copies it from a guide. 
Though, Alice wonders why \topos{} did not generate such a rule. 
She becomes skeptical about her decision and wants to double check. 
Asking \fffuu{} about the potential packet flows once a connection is initiated, \fffuu{} confirms that there are currently no limitations at all, not even for HTTP. 
She compares this to the stateful implementation computed by \topos{}, shown in Figure~\ref{fig:dynamicdocker:topos:dockermynet4.stateful}. 
The dashed orange line indicates a flow with stateful semantics, \ie packets may flow in both directions once the connection was initiated by the web application. 
She realizes that \topos{} takes great care to enforce unidirectional information flow to the log server. 
This is due to the information sink security invariant specified in the requirements. 
Alice knows from recent news that a badly protected log server may leak information which may lead to the compromise of all her machines~\cite{rhel2016logpwn}. 
Therefore, Alice restricts her \verb~ESTABLISHED~ rule to ssh. 
She uses the \texttt{multiport} module which conveniently allows matching on either source port or destination port in one rule. 
She makes the following final adjustment to her ruleset. 

\smallskip

\begin{minipage}{.95\linewidth}
	\footnotesize
	\begin{Verbatim}[commandchars=\\\{\},codes={\catcode`$=3\catcode`^=7}]
 -A CUSTOM
 -A CUSTOM -d 193.99.144.80/32 -m recent $\hfill\hookleftarrow$
     --set --name rateheise --rsource
 -A CUSTOM -d 193.99.144.80/32 -m recent $\hfill\hookleftarrow$
     --update --seconds 60 --hitcount 3 $\hfill\hookleftarrow$
     --name rateheise --rsource -j DROP
\diffdel{--A CUSTOM -p tcp -m state --state ESTABLISHED}$\hfill\hookleftarrow$
\diffdel{     -j ACCEPT}
\diffadd{+-A CUSTOM -p tcp -m state --state ESTABLISHED}$\hfill\hookleftarrow$
\diffadd{     -m multiport --ports 22 -j ACCEPT}
 -A CUSTOM -p tcp -m tcp --dport 22 -j ACCEPT
 COMMIT
	\end{Verbatim}
\end{minipage}

\medskip

Alice runs one final verification of the implemented policy with \fffuu{}, shown in Figure~\ref{fig:dynamicdocker:fffuu:dockermynet.topos4.1.state}. 
This time, she also includes the stateful flows. 
\fffuu{} identifies only one stateful flow, visualized with an orange dashed line. 
The direction of the stateful flow is the other way round compared to Figure~\ref{fig:dynamicdocker:topos:dockermynet4.stateful}. 
This is merely an artifact of the visualization, a stateful flow is essentially bidirectional once it is established. 
Otherwise, Figure~\ref{fig:dynamicdocker:topos:dockermynet4.stateful} and Figure~\ref{fig:dynamicdocker:fffuu:dockermynet.topos4.1.state} show isomorphic graphs. 
This verifies that Alice's firewall rules are correct.\footnote{For HTTP. Alice disregards ssh.}

\section{Related Docker Work}
\label{sec:relateddockerwork}
Tools to improve firewall management for Docker hosts exist~\cite{github2016dockerfw,github2016dfwfw}. 
Docker-fw~\cite{github2016dockerfw} is a convenient iptables wrapper with Docker-specific features, such as retrieving the IP address of a container using the Docker API. 
It currently only supports the default Docker bridge, but not custom networks. 
DFWFW~\cite{github2016dfwfw} is also a convenient tool to manage the iptables firewall on a Docker host. 
It runs as daemon and can apply changes dynamically if the Docker setup is modified, \eg if new containers are instantiated.

Both tools provide features that could help to make the management process with \topos{} and \fffuu{} more convenient. 
At the moment, \topos{} generates raw iptables rules but leaves the actual IP addresses to be set by the user, \eg $\mathit{\$WebFrnt\ippostfix{}}$ in Figure~\ref{fig:dynamicdocker:topos:dockermynet.topos4}. 
To further automate the setup, \topos{} could generate Docker-fw~\cite{github2016dockerfw} rules which automatically resolve the correct IP address. 
To further automate firewall management, \topos{} could directly generate DFWFW~\cite{github2016dfwfw} configurations. 
This would mean that no manual configuration is required any longer if multiple instances of the same container are spawned. 

Alice tests \topos{} together with DFWFW. 
For this, Alice simply adapts the \topos{} serialization step to generate rules in the DFWFW configuration format. 
Since DFWFW is also built to primarily support whitelisting, the translation is straightforward. 
A rule in this format first matches on the Docker network, then it allows specifying the source and destination container, an arbitrary string which will be added to the iptables match expression, and finally the iptables action. 
The match on the container names permits the use of Perl regular expressions. 
To allow dynamic spawning of multiple instances of a container, Alice writes a regex which matches on the container name and any trailing number, \eg \texttt{webfrnt}, \texttt{webfrnt1}, \texttt{webfrnt-1}, \texttt{webfrnt200}. 
The beginning of the configuration file looks as follows: 

\smallskip

\begin{minipage}{.95\linewidth}
	\footnotesize
	\begin{Verbatim}[commandchars=\\\{\},codes={\catcode`$=3\catcode`^=7}]
\{
  "container_to_container": \{
  "rules": [
    \{
       "network": "alicewebappnet",
       "src_container": "Name =\textasciitilde \textasciicircum{}webfrnt-?\textbackslash{}\textbackslash{}d*\$",
       "dst_container": "Name =\textasciitilde \textasciicircum{}webfrnt-?\textbackslash{}\textbackslash{}d*\$",
       "filter": "",
       "action": "ACCEPT"
     \}, 
     \{
       "network": "alicewebappnet",
       "src_container": "Name =\textasciitilde \textasciicircum{}webfrnt-?\textbackslash{}\textbackslash{}d*\$",
       "dst_container": "Name =\textasciitilde \textasciicircum{}log-?\textbackslash{}\textbackslash{}d*\$",
       "filter": "",
       "action": "ACCEPT"
     \}, 
       $\dots$
	\end{Verbatim}
\end{minipage}

\medskip

Disregarding the JSON formatting, it is similar to Figure~\ref{fig:dynamicdocker:topos:dockermynet.topos4}, but only the first two rules are shown. 
Alice tests that all containers have the necessary connectivity with this setting. 
Alice also tests that the firewall gets dynamically updated once she instantiates new containers and that the most obvious attempts to subvert the security policy are successfully blocked. 
She also verifies the generated iptables rules with \fffuu{}. 
This reveals that the overall setting is indeed good, but two open issues exist: 
First, Internet connectivity is again unconstrained and the stateless semantics are not enforced correctly, \ie once a connection with the log server is established, bidirectional communication is permitted. 
Alice leaves these engineering issues of fine tuning the DFWFW configuration to future work.

Ultimately, Alice already enhances the state-of-the-art of Docker container management by combining a dynamic Docker firewall framework with the static policy management tool \topos{}. 
While the Docker firewall deploys the security policy, \topos{} generates it. 
This combination lifts \topos{} to dynamic contexts since it allows dynamic spawning and deletion of containers at runtime while still providing strong guarantees about the enforced security requirements. 
In addition, \fffuu{} can verify the correctness of potentially modified firewall rules according to the policy at runtime.

\section{Survey of Related Academic Work}
\label{sec:sdnnfv:related}

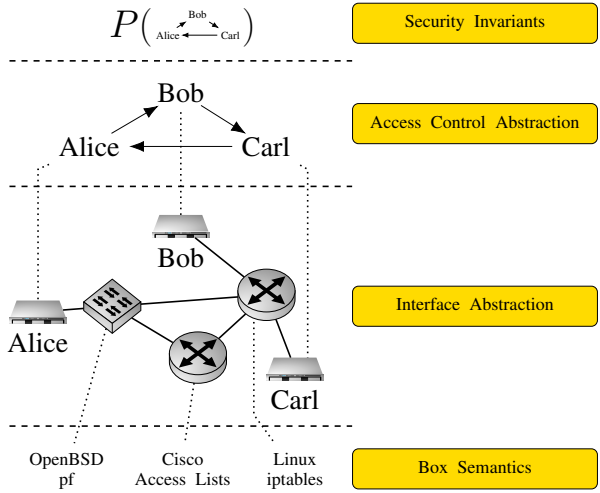
\begin{figure}[!htb]
	\centering%
\resizebox{0.9\linewidth}{!}{%
\Large
\begin{tikzpicture}
\node[text width=0.3\linewidth, align=center] (iheardyoulikerecursion) at (0,1.5) {
			\resizebox{.99\linewidth}{!}{
				\scalebox{2}{\Huge{$P($}}\Large\begin{tikzpicture} [baseline=-3ex]
				\node[anchor=south] (bi) at (0,0) {Bob};
				\node[anchor=east] (ai) at (-.9,-.6) {Alice};
				\node[anchor=west] (ci) at (+.9,-.6) {Carl};
				
				\draw[myptr] (ai) to (bi);
				\draw[myptr] (bi) to (ci);
				\draw[myptr,shorten >=1ex] (ci) to (ai.east); %
				\end{tikzpicture}\scalebox{2}{\Huge{$)$}}%
			}
		};

\def\sbordery{+.9}
\draw [thick,dashed] (-3,\sbordery)--(3,\sbordery);
\node[MyRoundedBox, fill=TUMYellow, anchor=south west, align=center, text width=8em] at ($(3.,1.5)!0.5!(3.,\sbordery)$) {\small{\strut{}Security Invariants}};

\node[anchor=south] (b) at (0,0) {Bob};
\node[anchor=east] (a) at (-.9,-.6) {Alice};
\node[anchor=west] (c) at (+.9,-.6) {Carl};

\draw[myptr,shorten >=-1ex,shorten <=-1ex] (a) to (b);
\draw[myptr,shorten >=-1ex,shorten <=-1ex] (b) to (c);
\draw[myptr] (c) to (a);

\def\abordery{-1.3}
\draw [thick,dashed] (-3,\abordery)--(3,\abordery);
\node[MyRoundedBox, fill=TUMYellow, anchor=west, align=center, text width=8em] at ($(3.,\sbordery)!0.5!(3.,\abordery)$) {\small{\strut{}Access Control Abstraction}};

\node[server,label=below:Alice] (ra) at (-2.5,-3.5) {};
\node[server,label=below:Bob] (rb) at (0,-2) {};
\node[server,label=below:Carl] (rc) at (2.,-4.5) {};
\node[switch] (aux1) at (-1.2,-3.4) {};
\node[router] (aux2) at (1.5,-3.2) {};
\node[router] (aux3) at (.3,-4.3) {};

\draw[thick] (ra) to (aux1);
\draw[thick] (rb) to (aux2);
\draw[thick] (rc) to (aux2);
\draw[thick] (aux1) to (aux2);
\draw[thick] (aux2) to (aux3);
\draw[thick] (aux1) to (aux3);

\draw [dotted,thick] (a)--(-2.5,-1)--(ra);
\draw [dotted,thick] (b)--(rb);
\draw [dotted,thick] (c)--($(2.,-1)+(1ex,0)$)--($(rc.north)+(1ex,0)$);

\def\ibordery{-5.5}
\draw [thick,dashed] (-3,\ibordery)--(3,\ibordery);
\node[MyRoundedBox, fill=TUMYellow, anchor=west, align=center, text width=8em] at ($(3.,\abordery)!0.5!(3.,\ibordery)$) {\small{\strut{}Interface Abstraction}};

\node[anchor=north] (l1) at (-2.,-5.8) {\small{\parbox[c]{5em}{\centering OpenBSD pf}}};
\draw [dotted,thick] (aux1)--(l1);

\node[anchor=north, align=center, text width=5em] (l3) at (0,-5.8) {\small{\parbox[c]{8em}{\centering Cisco Access~Lists}}};
\draw [dotted,thick] (aux3)--(l3);

\node[anchor=north] (l2) at (2.,-5.8) {\small{\parbox[c]{5em}{\centering Linux iptables}}};
\draw [dotted,thick] ($(aux2.south)+(-1ex,0)$)--($(1.5,-5.3)+(-1ex,0)$)--(l2);

\def\bbordery{-7}
\node[MyRoundedBox, fill=TUMYellow, anchor=west, align=center, text width=8em] at ($(3.,\ibordery)!0.5!(3.,\bbordery)$) {\small{\strut{}Box Semantics}};

\end{tikzpicture}}%
		\caption{Four Layer Abstractions}%
		\label{fig:4layerabstraction}%
\end{figure}

We first define four management abstraction layers to subsequently classify related work in the field of network management and security, not limited to Docker.
Each abstraction layer is responsible for an individual problem domain. 
We illustrate the four layers in Figure~\ref{fig:4layerabstraction}. 
The layers have well-defined interfaces, thus, it is possible to combine solutions of individual problems. 

We propose the following four layers of abstraction.
\begin{description}%
	\item[Security Invariants]
		Defines the high-level security goals.
		Representable as predicates.
		For example, Figure~\ref{fig:runexsecinvars}.
	\item[Access Control Abstraction]
		Defines the allowed accesses between policy entities.
		Representable as global access control matrix.
		For example, Figure~\ref{fig:secpol}.
	\item[Interface Abstraction]
		Defines a model of the complete network topology.
		Representable as a graph, packets are forwarded between the network entity's interfaces.
	\item[Box Semantics]
		Describes the semantics (\ie behavior) of individual network boxes.
		Usually, the semantics are vendor-specific (\eg iptables, Cisco ACLs, Snort IDS). 
\end{description}
The main difference between the interface abstraction and the box semantics is that the latter describes the behavior of only one network entity, whereas the former describes the interconnection of many, possibly different, network boxes. 
In our story, both coincide since only one Docker host is considered.
By separating the box semantics from the interface and access control abstraction, the low level implementation of the enforcement device can be exchanged, if the new devices can provide equal semantics.
Only the generation of the target configuration must be adapted.
For example, \topos{} can also used to generate an OpenFlow configuration \cite{diekmann2015topos}.

In Figure~\ref{fig:relatedworkabstractionlayers}, we summarize how related work bridges the abstraction layers.
With regard to the abstractions, work may be horizontal or vertical: 
Vertical work bridges abstraction layers, for example, translating security invariants (Figure~\ref{fig:runexsecinvars}) to the access control abstraction (Figure~\ref{fig:secpol}) is a vertical step. 
Horizontal work adds features or conducts safety checks on the same abstraction level, for example, augmenting the directed policy (Figure~\ref{fig:secpol}) to a stateful policy (Figure~\ref{fig:statefulpol}) is a horizontal step. 
A direct arrow from the access control abstraction to the box semantics (and vice versa) means that the solution only applies to a single enforcement box. 
Solutions such as Firmato and Fireman achieve more and are thus listed multiple times.

\newcommand{\AccessControlVerifySinvars}[0]{VALID~\cite{bleikertz2011VALID}; \topos{}\phantom{;}}
\newcommand{\SinvarsToAccessControl}[0]{Zhao~\etal\cite{zhao2011policyremanet}; \mbox{\topos{} step \emph{B}}}

\newcommand{\boxSemanticsToAccessControlAbstraction}[0]{Fireman~\cite{fireman2006}; ITVal~\cite{marmorstein2006firewall}; \mbox{\fffuu{}\phantom{;}}}
\newcommand{\InterfaceAbstractionToAccessControl}[0]{Fireman~\cite{fireman2006}; HSA~\cite{kazemian2012HSA}; Anteater~\cite{Mai2011anteater}; \mbox{ConfigChecker}~\cite{alshaer2009configchecker}; VeriFlow~\cite{khurshid2013veriflow}} %
\newcommand{\InterfaceAbstractionMAPSAccessControl}[0]{Xie~\cite{xie2005static}; Lopes~\cite{lopes2013msrnetworkverificationprogram}}
\newcommand{\BoxSemanticsMAPSInterfaceAbstraction}[0]{HSA~\cite{kazemian2012HSA}; Anteater~\cite{Mai2011anteater}; Config\-Checker~\cite{alshaer2009configchecker}}
\newcommand{\AccessControlToInterfaceAbstraction}[0]{one big\phantom{;} switch~\cite{monsanto2013composingonebigswitch}; Firmato~\cite{bartal1999firmato}; FLIP~\cite{ZhangAlShaer2007flip}; \mbox{FortNOX}~\cite{Porras2012FortNOX}; Merlin~\cite{soule2014merlin}; Kinetic~\cite{kinetic2015}; \mbox{PBM~\cite{dinesh2002policybased};} \mbox{NetKAT~\cite{icfp2015smolkanetkatcompiler}\phantom{;}}}
\newcommand{\AccessControlToBoxSemantics}[0]{Firmato~\cite{bartal1999firmato}; FLIP~\cite{ZhangAlShaer2007flip}; NetKAT~\cite{icfp2015smolkanetkatcompiler}; Mignis~\cite{mignis2014};\newline Or-BAC~\cite{orbacnetwork04}; \topos{} step \emph{C}+\emph{D}}
\newcommand{\InterfaceAbstractionToBoxSemantics}[0]{RCP~\cite{Caesar2005rcp}; \mbox{OpenFlow}~\cite{mckeown2008openflow}; Merlin~\cite{soule2014merlin}; \mbox{optimized one}\phantom{;} \mbox{big switch}~\cite{Kang2013onebigswitchabstraction}; NetKAT~\cite{icfp2015smolkanetkatcompiler}; VeriFlow~\cite{khurshid2013veriflow}; FML~\cite{hinrichs2009practical}\phantom{;}}
\newcommand{\BoxSemantics}[0]{\mbox{Iptables Semantics}~\cite{diekmann2015fm}} %

\begin{figure}[!htb]
\centering
    \resizebox{0.99\linewidth}{!}{%
 	\begin{tikzpicture}
	 \node [MyRoundedBox, fill=TUMYellow, text width=13em](abs1) at (0,0) {\strut{}Security Invariants};
	 \node [MyRoundedBox, fill=TUMYellow, text width=13em](abs2) at (0,-2) {\strut{}Access Control Abstraction};
	 \node [MyRoundedBox, fill=TUMYellow, text width=13em](abs3) at (0,-8) {\strut{}Interface Abstraction};
	 \node [MyRoundedBox, fill=TUMYellow, text width=13em](abs4) at (0,-14) {\strut{}Box Semantics};
	 
	 \draw [myptr] ($(abs1.south) + (+3em,0)$)--($(abs2.north) + (+3em,0)$);
	 \node [anchor=west, align=left, text width=12em] at ($(abs1)!0.5!(abs2) + (+3.2em,0)$) {\SinvarsToAccessControl};
	 
	 \draw [myptr,dashed] ($(abs2.north) + (-3em,0)$)--($(abs1.south) + (-3em,0)$);
	 \node [anchor=east, align=right, text width=7em] at ($(abs1)!0.5!(abs2) + (-3.2em,0)$) {\AccessControlVerifySinvars};
	 
	 \draw [myptr] ($(abs2.east) + (0,-0.8ex)$)--($(abs2.east) + (+6em,-0.8ex)$)--($(abs3.east) + (+6em,+0.8ex)$)--($(abs3.east) + (0,+0.8ex)$);
	 \node [anchor=east, align=right, text width=7em] at ($(abs2)!0.5!(abs3) + (+12.8em,0)$) {\AccessControlToInterfaceAbstraction};

	 \draw [myptr,dotted] ($(abs3.north) + (-2em,0)$)--($(abs2.south) + (-2em,0)$);
	 \node [anchor=west, align=left, text width=7em] at ($(abs3)!0.5!(abs2) + (-1.8em,0)$) {\InterfaceAbstractionMAPSAccessControl};
	 
	 \draw [myptr] ($(abs3.east) + (0,-0.8ex)$)--($(abs3.east) + (+6em,-0.8ex)$)--($(abs4.east) + (+6em,+0.8ex)$)--($(abs4.east) + (0,+0.8ex)$);
	 \node [anchor=east, align=right, text width=7em] at ($(abs3)!0.5!(abs4) + (+12.8em,0)$) {\InterfaceAbstractionToBoxSemantics};
	 \draw [myptr] ($(abs2.east) + (0,+0.8ex)$)--($(abs2.east) + (+6.8em,+0.8ex)$)--($(abs4.east) + (+6.8em,-0.8ex)$)--($(abs4.east) + (0,-0.8ex)$);
	 \node [anchor=west, align=left, text width=7em] at ($(abs2)!0.5!(abs4) + (+14.0em,0)$) {\AccessControlToBoxSemantics};

	 \draw [myptr,dotted] ($(abs4.north) + (-2em,0)$)--($(abs3.south) + (-2em,0)$);
	 \node [anchor=west, align=left, text width=7em] at ($(abs4)!0.5!(abs3) + (-1.8em,0)$) {\BoxSemanticsMAPSInterfaceAbstraction};	 
	 
	 \draw [myptr,dashed] ($(abs3.west) + (0,+0.8ex)$)--($(abs3.west) + (-6em,+0.8ex)$)--($(abs2.west) + (-6em,-0.8ex)$)--($(abs2.west) + (0,-0.8ex)$);
	 \node [anchor=west, align=left, text width=9em] at ($(abs2)!0.5!(abs3) + (-12.9em,0)$) {\InterfaceAbstractionToAccessControl};

	 \draw [myptr,dashed] ($(abs4.west) + (0,-0.8ex)$)--($(abs4.west) + (-0.8em,-0.8ex)$)--($(abs4.west) + (-0.8em,+5)$)--($(abs4.west) + (-6.8em,+5)$)--($(abs2.west) + (-6.8em,+0.8ex)$)--($(abs2.west) + (0,+0.8ex)$);
	 \node [anchor=east, align=right, text width=6em] at ($(abs2)!0.5!(abs4) + (-14.0em,0)$) {\boxSemanticsToAccessControlAbstraction};

	 \draw [myptr] ($(abs4.south) + (-3em,0)$)--($(abs4.south) + (-3em,-3ex)$)--($(abs4.south) + (+3em,+-3ex)$)--($(abs4.south) + (+3em,0)$);
	 \node [anchor=north, align=center, text width=10em] at ($(abs4.south)+ (0,-3ex)$) {\BoxSemantics};

	 \draw[myptr] (-3,-12) to node[above,sloped]{translates} (-3,-14);
	 \draw[myptrrev,dotted] (-2.25,-12) to node[above,sloped]{maps} (-2.25,-14);
	 \draw[myptrrev,dashed] (-1.5,-12) to node[above,sloped]{verifies} (-1.5,-14);
	 \end{tikzpicture}
	}
		\caption{Four Layer Abstraction in Related Work}
		\label{fig:relatedworkabstractionlayers}
\end{figure}
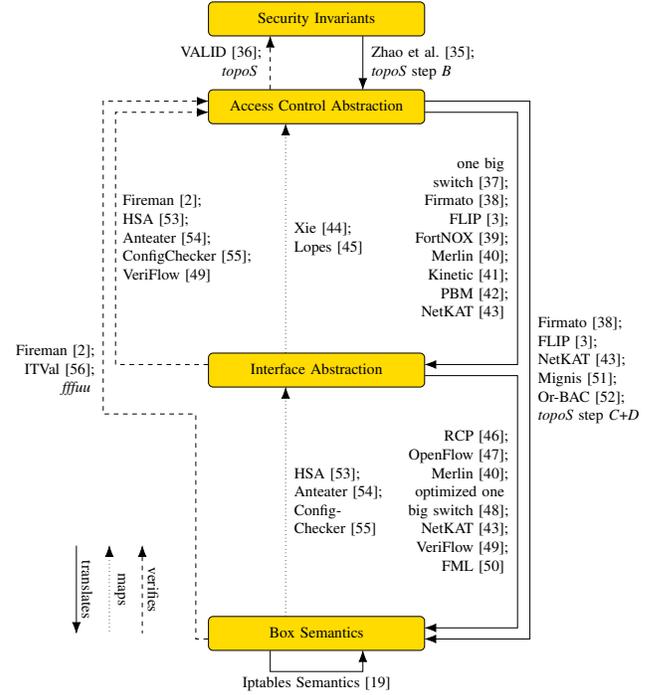

\paragraph*{Security Invariants $\rightarrow$ Access Control Abstraction}

Zhao \etal\cite{zhao2011policyremanet} present a policy refinement framework for network services. 
They state the need to express security requirements in high-level terms and present a high-level logic language to encode them. 
For this language, an automated translation procedure to the access control abstraction is presented. 
Also, low-level policies which can be enforced directly by certain security mechanisms are presented (not shown in Fig.\ \ref{fig:relatedworkabstractionlayers} since their current prototype implementation only supports one single policy rule). 
The logic-based, abstract policy language roughly corresponds to first-order logic with set theory, orderings, and relations over an UML class diagram. 
It allows to express almost unrestricted statements. 
The downside for a user of such a language may be that it is hard to specify and hard to follow. 
The system does not provide a reader with feedback about whether the encoded high-level security requirement actually corresponds to the policy author's intention. 
In contrast, \topos{} comes with a library of pre-specified logic formulas (security invariant templates). 
A policy author only needs to assign attributes, but is not required to manually encode requirements as formulas. 
In addition, \topos{} gives immediate feedback to a policy author by visualizing the resulting access control policy as graph. 
This enables a policy author to verify the requirement specification and perform \textit{``What-if?''} analyses. 

\begin{figure}[h!tb]
	\centering
  		\includegraphics[width=0.8\linewidth]{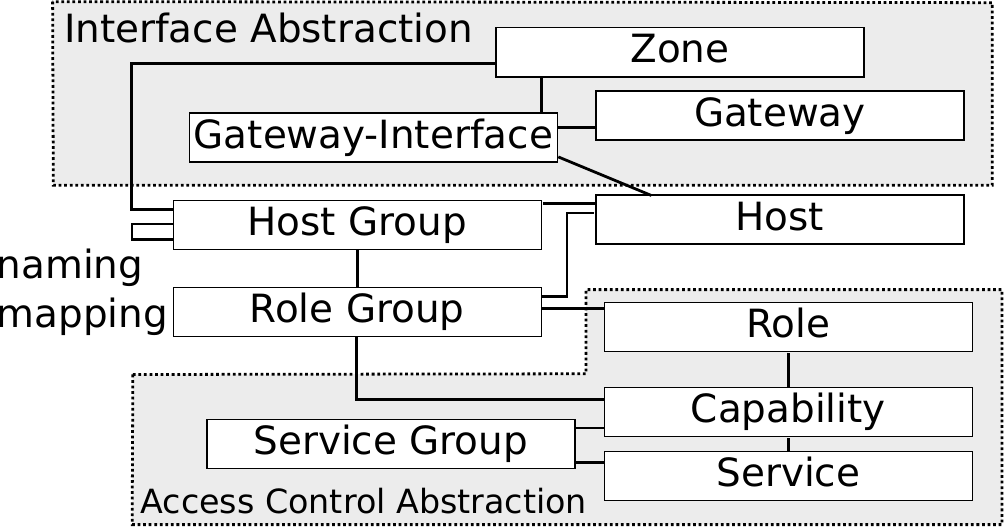}
  		\vskip-5pt
		\caption{Firmato ERM}
		\label{fig:firmatoerm}
\end{figure}

\paragraph*{Access Control Abstraction $\dashrightarrow$ Security Invariants}
Bleikertz and Gro{\ss} also state the need to express security requirements in a high-level language and verify policies against them. 
They propose VALID~\cite{bleikertz2011VALID} which is a specification language to express security invariants in cloud infrastructures. 
VALID has aspects in common with \topos{}, for example, it requires a network's connectivity and information flow structure in the format of a graph as input and allows specifying predicates over it. 
The language is formally defined in the AVISPA Intermediate Format. 
It can verify that a given network topology conforms to specified high-level goals, but it cannot translate these goals to a network topology nor give feedback about the meaning of the specified goals. 
\paragraph*{Access Control Abstraction $\rightarrow$ Interface Abstraction}
From the point of view of our fictional story, Firmato~\cite{bartal1999firmato} is probably the work most closely related to \topos{}. 
It defines an entity relationship model %
to structure network management and compile firewall rules from it, illustrated in Figure~\ref{fig:firmatoerm}. 
Firmato focuses on roles, which in our model correspond to container names. 
A role has positive capabilities and is related to other roles, which can be used to derive an access control matrix. 
Zones, Gateway-Interfaces, and Gateways define the network topology, which corresponds to the interface abstraction. 
As illustrated in Figure~\ref{fig:firmatoerm}, the abstraction layers identified in this work can also be identified in Firmato's model. 
The Host Groups, Role Groups, and Hosts definitions provide a mapping from policy entities to network entities, which is Firmato's approach to the naming problem. 
With close correspondence in the underlying concepts to Firmato, Cuppens \etal\cite{Cuppens2005orbacxmlfirewall} propose a firewall configuration language based on Or-BAC~\cite{orbac}. 
Similar to Firmato (with more support for negative capabilities) is FLIP~\cite{ZhangAlShaer2007flip}, which is a high-level language with focus on \emph{service} management (\eg allow/deny HTTP). 
Essentially, both FLIP and Firmato enhance the access control abstraction horizontally by including layer four port management and traverse it vertically by serializing to firewall rules.

FML~\cite{hinrichs2009practical} is a flow-based declarative language to define, among others, access control policies in a DATALOG-like language. 
Comparably to our directed (stateless) policy, FML operates on unidirectional network flows. 
FML solves the naming problem by assuming that all entities are authenticated with IEEE 802.1X~\cite{IEEE8021X}.

Policy-based management (PBM)~\cite{dinesh2002policybased} was introduced to simplify network administration. 
Similar to our work, it proposes different levels of abstraction and describes how to translate between them. 
Policy-based management defines a generic information model~\cite{rfc3060,rfc3460} which is not limited to access control, 
however, we focus our discussion solely on access control and security. 
In a central policy repository, global policy rules are stored. 
Policy decision points retrieve these rules and interpret them. 
Using our terminology, this step translates the access control abstraction to the interface abstraction. 
A policy decision point forwards decisions to policy enforcement points, implementing the translation from the interface abstraction to box semantics. 
This last step may be very device-specific~\cite{hinrichs99pbmcisco} and is not the core focus of PBM. 
While PBM was built on the idea of specifying business-level abstractions in terms of requirements \cite{dinesh2002policybased}, 
the IETF specified a rule-based policy repository~\cite{rfc3060,rfc3460}, which restricts storing high-level requirements that cannot easily be expressed as rules. 
In strong contrast to \topos{}, which focuses on specifying higher-level requirements (\ie security invariants), the IETF Policy Framework working group focused on the specification of lower-level policies~\cite[\S\hairspace{}2.1. Policy Scope]{rfc3060}. 
This can also be witnessed in many languages which were developed over the years~\cite{Han2012surveypbm} since, in particular when it comes to security, they usually only provide access control abstractions.

The ``One Big Switch'' Abstraction~\cite{monsanto2013composingonebigswitch,Kang2013onebigswitchabstraction,reich2013modular} allows to manage a network as if it were only one, central big switch. 
This effectively allows solutions which only support to manage one device to be applied to a complete network, consisting of multiple switches. 
With regard to Figure~\ref{fig:relatedworkabstractionlayers}, any solution which supports translating the access control abstraction to box semantics can also be applied to translate from the access control abstraction to the interface abstraction by translating to the ``One Big Switch''.

NetKAT~\cite{anderson2014netKATsemantics,icfp2015smolkanetkatcompiler} is a SDN programming language with well-defined semantics. 
It features an efficient compiler for local, global, and virtual programs to flow table entries~\cite{icfp2015smolkanetkatcompiler}. 
Among others, it allows implementing the ``One Big Switch'' Abstraction~\cite{icfp2015smolkanetkatcompiler}. 

\paragraph*{Access Control Abstraction $\rightarrow$ Box Semantics}
Mignis~\cite{mignis2014} proposes a declarative language to manage Netfilter/iptables firewalls. 
They focus on packet filtering and NAT. 
Their policy language is restrictive to avoid the usual policy conflicts. 
In particular, it only allows to write blacklisting-style policies, additionally NAT and filtering cannot arbitrarily be mixed to provide consistent policies.\footnote{The assumptions for the translation impose many restrictions \cite[\S\hairspace{}5]{mignis2014}.} 
Apart from \fffuu{}, it is the only work we are aware of which provides a formal semantics of an iptables firewall. 
They describe a semantics of the packet filtering and NAT behavior of iptables. %
Their semantics only models the aspects of iptables which are required for their policy language. 
While it supports NAT (which the semantics of \fffuu{} does not\footnote{Note that Docker utilizes iptables NAT features, but not in the \texttt{filter} table, where usually all access control decisions are made. 
This is the reason that \fffuu{} (though it does not support NAT) could be used to analyze Docker firewall filtering.}), it does not support user-defined chains or arbitrary match conditions.\footnote{Mignis supports match conditions but imposes additional assumptions on them to ensure that they are consistent with NAT. For example, the \texttt{iprange} module is not allowed. Without inspecting all matching modules manually, there is no generic way to assure whether a filter condition is compatible with Mignis.} 
Though supporting NAT and stateful filtering, advanced iptables features such as \texttt{NOTRACK} or \texttt{connmark} are not considered.

Craven \etal\cite{craven2011policyrefinement} present a generalized (not network-specific) process to translate access control policies, enhanced with several aspects, to enforceable device-specific policies; the implementation requires a model repository of box semantics and their interplay.
Pahl~\cite{Pahl2015IM} delivers a data-centric, network-specific approach for managing and implementing such a repository, further focusing on things. %

\paragraph*{Interface Abstraction $\dashrightarrow$ Access Control Abstraction }
As illustrated in Figure~\ref{fig:relatedworkabstractionlayers},
Fireman~\cite{fireman2006} is a counterpart to Firmato.
It verifies firewall rules against a global access policy.
In addition, Fireman provides verification on the same horizontal layer (\ie finding shadowed rules or inter-firewall conflicts, which do not affect the resulting end-to-end connectivity but are still most likely an implementation error).
Abstracting to its uses, one may call rcc~\cite{Feamster2005rcc} the fireman for BGP. 
\fffuu{} is unique as it not only verifies rules, but also translates them back to the access control abstraction. 

Header Space Analysis \allowbreak{}\mbox{(HSA)~\cite{kazemian2012HSA}}, Anteater \cite{Mai2011anteater}, and Config\-Checker~\cite{alshaer2009configchecker} verify several horizontal safety properties on the interface abstraction, such as absence of forwarding loops.
By analyzing reachability~\cite{xie2005static,lopes2013msrnetworkverificationprogram,Mai2011anteater,alshaer2009configchecker,kazemian2012HSA}, horizontal consistency of the interface abstraction with an access control matrix can also be verified. %
Verification of incremental changes to the interface abstraction can be done in real-time with VeriFlow~\cite{khurshid2013veriflow} and NetPlumber~\cite{kazemian2013realtimehsa}, the former can also prevent installation of violating rules. 
These models of the interface abstraction have many commonalities: %
The network boxes in all models are stateless and the network topology is a graph, connecting the entity's interfaces.
A function models packet traversal at a network box. %
These models could be considered as a giant (extended) finite state machine (FSM), where the state of a packet is an (interface$\times$packet) pair and the network topology and forwarding function represent the state transition function~\cite{lopes2013msrnetworkverificationprogram,zhang2012erificationswitching}.
In contrast to arbitrary state machines, it is believed that those derived from networks are comparatively well-behaved~\cite{zhang2012erificationswitching}.
Anteater~\cite{Mai2011anteater} differs in that interface information is implicit and packet modification is represented by relations over packet histories.

\paragraph*{Horizontal Enhancements}

Most analysis tools make simplifying assumptions about the underlying network boxes.
Diekmann \etal\cite{diekmann2015fm} present simplification of iptables firewalls. 
This makes complex real-world firewalls available for tools which were built with simplifying assumptions about rulesets. 
This horizontally simplifies the box semantics and \fffuu{} is built on top of these simplification procedures~\cite{diekmann2016networking}.

FortNOX~\cite{Porras2012FortNOX} horizontally enhances the access control abstraction as it assures that rules by security apps are not overwritten by other apps.
Technically, it hooks up at the access control/interface abstraction translation. %
Kinetic~\cite{kinetic2015,kineticwebsite} is an SDN language which lifts static policies (as constructed by \topos{}) to dynamic policies.
To accomplish this, an administrator can define a simple FSM which dynamically (triggered by network events) switches between static policies. 
In addition, the FSM can be verified with a model checker.

Features are horizontally added to the interface abstraction:
a routing policy allows specifying \emph{paths} of network traffic~\cite{Kang2013onebigswitchabstraction}. 
Merlin~\cite{soule2014merlin,soule2014merlinsconferenceversion} additionally supports bandwidth assignments and network function chaining. 
Both translate from a global policy to local enforcement and Merlin provides a feature-rich language for interface abstraction policies.
Conceptually similar (but with a completely different implementation), RCP~\cite{Caesar2005rcp} allows to logically centralize routing while remaining compatible with existing routers.

\section{Comparison to Academic State-of-the-Art}
\label{sec:relatedcomparison}
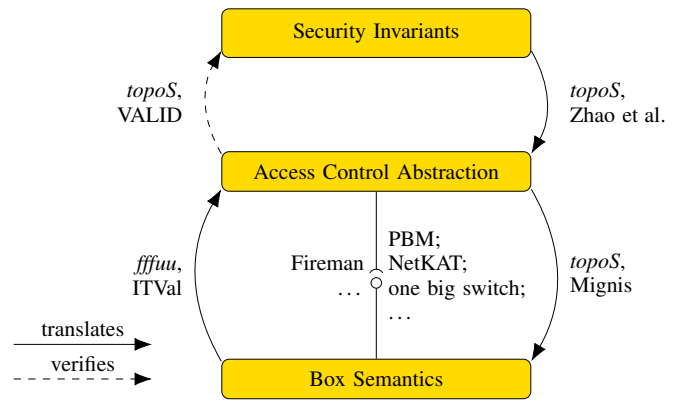
\begin{figure}[!htb]%
	\centering\small%
	\resizebox{0.99\linewidth}{!}{%
	\begin{tikzpicture}%
	\node [MyRoundedBox, fill=TUMYellow, text width=13em](abs1) at (0,-2) {\strut{}Security Invariants};
	\node [MyRoundedBox, fill=TUMYellow, text width=13em](abs2) at (0,-4) {Access Control Abstraction};
	\node [MyRoundedBox, fill=TUMYellow, text width=13em](abs4) at (0,-7) {Box Semantics};
	
	\draw [myptr] ($(abs1.east)+(0,-1.8ex)$) to[bend left] ($(abs2.east)+(0,+1.8ex)$);
	\node [anchor=west, align=left, text width=4em] at ($(abs1)!0.5!(abs2) + (+8.5em,0)$) {\topos{}, Zhao~et$\;$al.};
	
	\draw [myptr,dashed] ($(abs2.west)+(0,+1.8ex)$) to[bend left] ($(abs1.west)+(0,-1.8ex)$);
	\node [anchor=east, align=right, text width=3em] at ($(abs1)!0.5!(abs2) + (-8.5em,0)$) {\topos{}, VALID};
	
	\draw [myptr] ($(abs2.east) + (0,-1.8ex)$) to[bend left] ($(abs4.east) + (0,+1.8ex)$);
	\node [anchor=west, align=left, text width=4em] at ($(abs2)!0.5!(abs4) + (+8.5em,0)$) {\topos{}, \mbox{Mignis}};
	
	\draw [myptr] ($(abs4.west) + (0,+1.8ex)$) to[bend left] ($(abs2.west) + (0,-1.8ex)$);
	\node [anchor=east, align=right, text width=3em] at ($(abs2)!0.5!(abs4) + (-8.5em,0)$) {\fffuu{}, ITVal};
	
	\draw [-to,-(] ($(abs2.south)$)--($(abs4.north)!0.5!(abs2.south) + (0,.2ex)$);
	\draw [-to,o-] ($(abs4.north)!0.5!(abs2.south) + (0,-.2ex)$)--($(abs4.north)$);
	\node [anchor=west, align=left, text width=7em] at ($(abs2)!0.5!(abs4) + (+0.2em,0)$) {PBM;\newline{}\mbox{NetKAT;}\newline{}\mbox{one big switch;}\newline{}\dots};
	\node [anchor=east, align=right, text width=4em] at ($(abs2)!0.5!(abs4) + (-0.2em,0)$) {Fireman \dots};
	 
	 \draw[myptr] (-3,-6.5) to node[above,sloped]{translates} ++(2,0);
	 \draw[myptr,dashed] (-3,-7) to node[above,sloped]{verifies} ++(2,0);
	\end{tikzpicture}
	}
	\caption{Overview of Comparison to Related Work (cf.\ Fig.\ \ref{fig:relatedworkabstractionlayers})}
	\label{fig:dockerrelated}
\end{figure}

We are not aware of any academic related work about network access control management specifically for docker. 
To the best of our knowledge, \topos{} and \fffuu{} are the only academic works where applicability for a Docker environment has been specifically demonstrated. 
But Docker is merely an application example, both tools were not specifically designed for it but worked out of the box.\footnote{The power of formally verified code.} 
We broaden the scope and compare their underlying theory to the general state of the art of tools for helping network management and administration. 
The comparison is outlined in Figure~\ref{fig:dockerrelated}. 
It is aligned similar to Figure~\ref{fig:relatedworkabstractionlayers}, but we omit the Interface Abstraction. 
The solid single arrows visualize work which is able to translate between the corresponding abstractions. 
The dashed single arrow represents work which is only capable of verification: 
Given an access control policy and security invariants as input, their conformance can be verified. 
But one cannot be derived from the other. 
In general, a translation from an access control policy to security invariants is not possible without guessing a policy author's intention.

Our combination of \topos{} and \fffuu{} is the only compatible toolset which is able to bridge all levels of abstractions in both directions out of the box. 
The fictional story reveals that this back-and-forth is useful in several scenarios. 
The story also reveals features an administrator may wish for. 
We now compare \topos{} and \fffuu{} with related work specifically for the following use cases, considered in isolation. 

\paragraph*{Build Networks Based on a Security Requirement Specification}
Instead of writing a policy or low-level configuration by hand, the fictional story shows that it is useful to generate working network configurations directly from a scenario-specific security requirement specification. 
This is useful for the initial design and implementation, as well as for starting over in certain scenarios. 

VALID~\cite{bleikertz2011VALID} allows to express security requirements, but it cannot derive network configurations from them. 
Zhao \etal\cite{zhao2011policyremanet} also propose a framework which allows to express security requirements. 
In contrast to VALID, their framework additionally allows to derive working network configurations. 
The language proposed by Zhao \etal exposes a lot of formalism to the administrator and almost bears more resemblance to programming than it bears to a specification. 
Compared to this, \topos{} distinguishes strongly between templates and instantiating a template. 
While defining new templates also bears resemblance to programming and is only intended for expert users, the common operation to define a specification is by instantiating those templates, which only requires configurations and exposes very little formalism to the administrator.

\paragraph*{Allow Intervention and Low-Level Control for the Administrator}
The fictional story revealed several occasions where Alice wanted to fine-tune the low-level policy by hand. 
One example was the temporary, ad-hoc permission for ssh. 
Another example was the rate limiting to \url{heise.de}. 
It may also be imaginable that Alice needs to reorder some rules at some point for performance reasons. 
Both are true low-level operation which should not be achievable on a higher level of abstraction.\footnote{tautologically, higher levels of abstraction abstract over low-level details.} 

The Mignis~\cite{mignis2014} firewall configuration language allows to specify filtering policies. 
It gives the administrator the optional possibility to add arbitrary additional low-level iptables match conditions to the high-level rules. 
These additional match conditions may introduce soundness issues. 
The language does not permit the administrator to change the generated iptables rules directly, \eg reordering, restructuring, or ad-hoc changes without recompiling are not allowed. 
In contrast, \topos{} permits arbitrary changes to the generated iptables rules since \fffuu{} can be used to verify correctness of the changes.

\paragraph*{Detect Erosion and Drift of the Implemented Policy vs.\ the Specified Policy}
The terms \emph{erosion} and \emph{drift} are usually used for software architectures~\cite{Perry1992softwarearchitectures}. 
However, our fictional story shows that network security policies and the corresponding configurations also decay, become unmaintainable, and violations of the original requirements sneak in. 
In addition, being able to detect differences between a configuration and a specified policy is an important step towards understanding legacy configurations or verifying manual low-level changes, as discussed in the previous paragraph. 

To analyze the current policy enforced by an iptables ruleset, ITVal~\cite{marmorstein2005itval,marmorstein2006firewall} can be used. 
Similar to \fffuu{}, it allows to partition the complete IPv4 address range into equivalence classes. 
ITVal computes one set of equivalence classes jointly for all layer~4 ports. 
In contrast, \fffuu{} can only compute them for one selected port. 
It depends on the scenario which of the two approaches is more suitable: 
ITVal's overview is very helpful for a first, quick, overview of a firewall. 
\fffuu{}'s service-specific overview provides better granularity, once one knows which ports one is interested in. 
ITVal only supports IPv4 while \fffuu{} supports IPv4 and IPv6. 
In addition, ITVal is known to have bugs~\cite{diekmann2016networking} while \fffuu{} is formally proven correct~\cite{Iptables_Semantics-AFP}. 
Ironically, ITVal segfaults for some Docker rulesets of this article while \fffuu{} processes them without complaint. 

\paragraph*{Distributed Enforcement?}
Our work only focuses on one single, central enforcement device. 
However, having only one central firewall or only one central Docker host is not a satisfying scenario. 
This raises the question whether our work is useless for large installations or whether distributed enforcement is an orthogonal issue.
The Interface Abstraction---discussed in Section~\ref{sec:sdnnfv:related} and omitted in Figure~\ref{fig:dockerrelated}---corresponds to distributed enforcement. 
Figure~\ref{fig:relatedworkabstractionlayers} lists several related work which takes one centralized policy and enforces it in a distributed fashion. 
Therefore, distributed enforcement is an orthogonal issue and our tools can help to develop, verify, and maintain the centralized policy which is then enforced in a distributed fashion. 
For example, policy-based management~\cite{dinesh2002policybased} systems or the NetKAT compiler~\cite{icfp2015smolkanetkatcompiler} for SDN could be used. 
We explicitly visualize an interface boundary \tikz[baseline]{
\draw [thick,-to,-(] (0,.7ex)--($(.5,.7ex) + (-.2ex,0)$);
\draw [thick,-to,o-] ($(.5,.7ex) + (.2ex,0)$)--(1,.7ex);
} in Figure~\ref{fig:dockerrelated} to highlight that \topos{} produces an access control matrix, which is understood by many technologies for distributed enforcement. 
Therefore, \topos{} can be used as a module for access control within another system. 
In addition, algorithms for distributed firewall analysis (as supported by Fireman~\cite{fireman2006}) can also benefit from the pre-processing and simplification provided by \fffuu{}.

\section{Conclusion}
\label{sec:conclusion}
We presented our tools \topos{} and \fffuu{}, demonstrating design, management, and operations of network-level access control. %
In a fictional story about an operator in a Docker-based environment, we showed how this toolset helps both for the design of a setup and for daily operations. 
We demonstrated several situations in which our tools provide useful feedback, uncover bugs, and even help to migrate setups. 
The duality of \topos{} and \fffuu{} in combination with their common policy abstraction makes them a powerful combination and enhances the academic state-of-the-art.

The underlying theory of both tools is formally verified and their code is directly generated by Isabelle/HOL, providing strong correctness guarantees about their results. 
Having proven the correctness of the tools, no theorem prover is required at runtime; the code is correct for all inputs. 
Our tools are the first, jointly designed, formally machine-verified, open source, real-world-approved tools which bridge the gaps between high-level security requirements and low-level firewall behavior in both directions. 
Notably, they are not limited to Docker environments, but also applicable to different scenarios. 
This becomes explicit since they were never designed for Docker specifically, but worked flawlessly for the Docker scenarios. 
We surveyed related work, showed how our tools enhance the state-of-the-art, and how different tool may interact given common abstractions.

\bigskip
\noindent
\textbf{Availability: }
\noindent\topos{} and \fffuu{} can be obtained at %
\begin{center}
\url{https://github.com/diekmann/topoS}\\
\url{https://github.com/diekmann/Iptables_Semantics}
\end{center}

\noindent
A stable version of the theory files can also be obtained from the Archive of Formal Proofs (AFP)~\cite{Network_Security_Policy_Verification-AFP,IP_Addresses-AFP,Simple_Firewall-AFP,Iptables_Semantics-AFP,Routing-AFP}. 
AFP maintenance policy ensures that our formalization will keep working with future Isabelle releases. 
The raw data of the Docker story and many other iptables dumps can be found at  %
\begin{center}
\url{https://github.com/diekmann/net-network}
\end{center}
To the best of our knowledge, this is the largest, publicly-available collection of real-world iptables firewall rulesets. 
 
\section*{Acknowledgment}
The authors would like to thank the anonymous reviewers for their valuable feedback.
This work was supported by the German Federal Ministry of Education and Research under the projects SENDATE-PLANETS (16KIS0472) and DecADe (16KIS0538) and the German-French Academy for the  Industry of the Future.

\bibliographystyle{IEEEtran}
\bibliography{literature}
\end{document}